\documentclass[aip,jcp,reprint,superscriptaddress,footinbib,floatfix]{revtex4-2}

\usepackage{graphicx}
\usepackage{scalerel}
\usepackage[version=4]{mhchem}
\usepackage{latexsym}
\usepackage{textcomp}
\usepackage{amssymb}
\usepackage{amsbsy}
\usepackage{amsmath}
\usepackage{setspace}
\usepackage{epstopdf}
\usepackage{bm}
\usepackage{longtable}
\usepackage{subfigure}
\usepackage{epsfig}
\usepackage{dcolumn}
\usepackage{mathrsfs}
\usepackage[Euler]{upgreek}
\usepackage{multirow}
\usepackage{footnote}
\usepackage{stmaryrd}
\usepackage{tabulary}
\usepackage{color}
\usepackage[normalem]{ulem} 
\usepackage{tikz}
\usepackage{mathtools}
\usepackage{empheq}
\usepackage{cancel}
\usepackage{breqn}
\usepackage{xcolor}
\usepackage[most]{tcolorbox}
\usepackage{mathtools}
\usepackage{hyperref}

\usepackage{breqn}

\makeatletter
\let\cat@comma@active\@empty
\makeatother

\makeatletter 
\gdef\@ptsize{2}
\let\@currsize\normalsize 
\makeatother 
\usepackage{setspace} 
\singlespace

\begin{document}
\title{Dissipative response of driven bead-spring-dashpot chains}
\date{\today}
\author{R. Kailasham}
\email{rkailasham@iiti.ac.in}
\affiliation{Department of Chemical Engineering, Indian Institute of Technology Indore, Khandwa Road, Simrol, Madhya Pradesh -  453552, India}

\begin{abstract}
The work dissipated in pulling a polymer chain with internal friction is numerically calculated by considering a sequence of $N$ spring-dashpots tethered at one end and being pulled at the other using a harmonic trap via linear and symmetric protocols. 
The variation of the dissipation with the chain length, pulling trap stiffness, and the internal friction parameter are examined in detail for both the protocols. {While the dissipation increases with $N$ for chains without internal friction, the relationship between the dissipation and $N$ for chains with internal friction depends on the pulling trap stiffness. For chains with internal friction, the dissipation \textit{decreases} (increases) with $N$ as the the pulling trap stiffness is \textit{increased} (decreased), keeping all other parameters constant. Therefore, unlike in the case of a single-mode spring-dashpot ($N=1$) for which a simple relationship exists between the damping coefficient of the dashpot and the dissipated work [as shown in Phys. Rev. Res. \textbf{2}, 013331 (2020)], the same is not true for the general case of $N>1$ due to the stiffness-dependent dissipative response of bead-spring-dashpot chains.}
\end{abstract}

\maketitle

\section{\label{sec:intro} Introduction}

{The dynamics of conformational transitions in polymer molecules is governed not only by the solvent friction}, but also by intramolecular interactions whose effects are collectively referred to as internal friction or internal viscosity~\cite{kuhn1945,Booij1970,Manke1987,Hua1995,Hua1996,Kailasham2018,Mackay1992,Massa1971,Ansari1992,Jas2001,Qiu2004,Cellmer2008,Hagen2010385,Wensley2010,Soranno2012,DeSancho2014,Echeverria2014,Schulz20154565,Samanta2016,Soranno2017,Poirier2002,Murayama2007,Khatri20071825,Alexander-Katz2009,Einert2011,Ojala2014,Mondal2020,Nandagiri2020} (IV). The rheological consequences of internal friction~\cite{kuhn1945,Booij1970,Manke1987,Hua1995,Hua1996,Kailasham2018} include the appearance of a stress jump at the inception of shear flow~\cite{Mackay1992}, and a non-vanishing dynamic viscosity in the high frequency limit~\cite{Massa1971} of small amplitude oscillatory shear experiments. In the biophysical context, internal friction slows down the reconfiguration or folding time of the molecule~\cite{Ansari1992,Jas2001,Qiu2004,Cellmer2008,Hagen2010385,Wensley2010,Soranno2012,DeSancho2014,Echeverria2014,Schulz20154565,Samanta2016,Soranno2017}. While these two aspects of IV have received immense scrutiny, the mechanical or dissipative consequence of internal friction remains less thoroughly explored~\cite{Poirier2002,Murayama2007,Khatri20071825,Alexander-Katz2009,Einert2011,Ojala2014,Mondal2020,Nandagiri2020}. The work dissipated in the forced uncoiling of condensed DNA globules~\cite{Murayama2007,Ojala2014}, the viscoelastic response of single molecules of polysaccharides~\cite{Khatri20071825}, ciliary oscillations in microorganisms~\cite{Mondal2020}, are all modulated by the presence of a frictional mechanism that is not hydrodynamic in nature. Given the wide range of contexts in which internal friction modulates the mechanical response of molecules, it is therefore essential to quantify the work required in subjecting polymer chains with internal friction to various driving protocols. In this paper, we address this knowledge gap by analytically deriving the appropriate expressions for a coarse-grained polymer model.

\begin{figure*}[t]
\centering
\includegraphics[width=170mm]{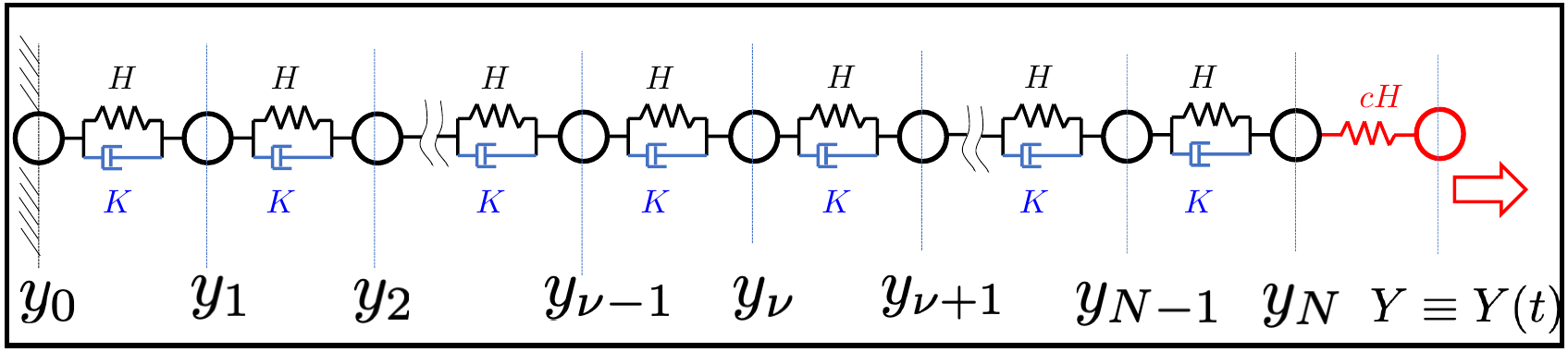}  
\caption{A polymer chain with $N+1$ beads, with one end ($y_0$) tethered and the other ($y_{N}$) connected to an optical/laser trap whose center $Y(t)$ moves according to a deterministic protocol. The spring-dashpot arrangement that connects adjacent beads is composed of a Hookean spring of stiffness $H$ and a dashpot with damping coefficient $K$. The stiffness of the steering spring is given by $cH$, with $c>0$. The motion of the $N$ beads numbered $y_1,y_2,\cdots,y_{N}$ is governed by eq.~(\ref{eq:coll_coord_eom}).}
\label{fig:bsp_chain_schem}
\end{figure*}

The concept of internal friction is an emergent phenomenon~\cite{rk2021}, and need not be invoked in an all-atom model of polymeric chains in which the interaction between various segments is specified completely in terms of bonded and non-bonded potentials~\cite{ja211494h,Schulz20154565}. The process of coarse-graining such a detailed model into a chemistry-agnostic bead-spring chain model requires the inclusion of an additional frictional term~\cite{degennes,Khatri20076770} that accounts for resistance to dihedral angle rotation~\cite{Fixman1988,Jas2001,Khatri20076770,Echeverria2014,DeSancho2014,Daldrop2018,Swiatek2024}. {This additional source of damping, or internal friction, is represented in coarse-grained models by the inclusion of a dashpot in parallel with each spring of a bead-spring-chain model~\cite{kuhn1945,Booij1970,Manke1987,Dasbach1992,Hua1995,kailasham2021rouse}. Internal friction or internal viscosity (IV) is a rate-dependent phenomenon that does not feature in the Hamiltonian of a coarse-grained micromechanical polymer model. Consequently, IV leaves the static equilibrium properties of the polymer (e.g., the probability distribution of the end-to-end distance~\cite{Kailasham2023}) unchanged. The signature of IV is therefore expected to be discernible from the dynamical properties of the polymer molecule, such as the timescale for loop-formation between two segments of a chain. Biophysicists have experimentally measured reconfiguration or folding time statistics of various molecules, and argue that the folding time has an internal friction component and a solvent-based component. By fitting experimental data to the Rouse model with internal friction (RIF), they have been able to extract the solvent- and IV-dependent contributions to the total reconfiguration time~\cite{Piana2015,Soranno2017,Ameseder2018}. This approach, however, does not yield a friction coefficient but rather a timescale.} 

{A survey of the literature indicates that single molecule force spectroscopic measurements~\cite{Harris2007,Murayama2007,Khatri20071825}, rather than reconfiguration time measurements, are needed to obtain the friction coefficient associated with polymer molecules. The idea is to perturb the polymer molecule away from equilibrium by stretching, and anticipate the work done in the process would contain the signature of internal friction~\cite{Alexander-Katz2009,Schulz20154565}. The total work expended in driving a system from an initial state to a final state has a reversible (free-energy) component and an irreversible (dissipative) component. Since IV is a dissipative phenomenon, it does not affect the free energy of the system. However, like any frictional mechanism, it is expected to affect the dissipation incurred in driving a system between two states.}

{Almost all previous theoretical studies examining the work statistics of polymers being pulled deal with the bead-spring chain (or Rouse) model~\cite{Dhar2005,Speck2005,Varghese2013}. An exception~\cite{Kailasham2020} is the simplest case of a dumbbell model with internal friction (two beads connected by a spring-dashpot arrangement) whose one bead is tethered and the other is pulled at a constant velocity. For this case, a direct connection has been established between the damping coefficient of the dashpot and the friction coefficient associated with pulling the trapped bead over a fixed distance at a constant velocity. This mechanical interpretation of internal friction allows for an unambiguous quantification of the phenomenon. Polymer chains are usually modeled as a chain of interconnected beads, rather than as a dumbbell with only two beads, to capture the multiple timescales associated with their dynamics~\cite{Ottinger1996,Bird1987b}. To the best of our knowledge, however, there have been no attempts so far to answer what the dissipation incurred in pulling a series of spring-dashpots is. Also, an accurate coarse-grained description of polymers with internal friction would require the knowledge of how the dissipation of a driven bead-spring-dashpot chain is related to the damping coefficient of a single dashpot. The present work addresses these two questions, by obtaining semi-analytical expressions for the dissipation under various protocols.}

{The work statistics of a Rouse chain whose one end is fixed and the other subjected to one-dimensional motion via pulling at a constant velocity or oscillation at a fixed frequency have been derived~\cite{Dhar2005,Speck2005,Varghese2013}. We extend the approach to a bead-spring-dashpot chain, and derive expressions for the work statistics under both the protocols. We discover, however, that the dependence of the dissipation on the number of spring-dashpots in the chain is highly non-trivial, and is dictated by the stiffness of the steering spring relative to that of a single spring in the chain. Essentially, a soft steering spring results in the dissipation increasing with $N$, while a stiff spring results in the dissipation decreasing with $N$ (termed ``anti-cooperative response" and explained later in the paper). This seemingly surprising, stiffness-dependent response precludes the establishment of a straightforward relationship between the dissipation incurred in pulling a chain to the damping coefficient of a single dashpot. A stiffness-dependent response to pulling has been exploited for the design of minimally-dissipative protocols~\cite{Blaber2022,Blaber2022a} for manipulating the positions of colloids moving in a corrugated external potential. The observation of a stiffness-dependent dissipative response in a coarse-grained model for polymers with internal friction is the central physical finding of this paper.}

This document is organized as follows. Section~\ref{sec:gov_eq} describes the bead-spring-dashpot chain model and derives the expression for the work dissipated in subjecting a tethered chain to an arbitrary pulling protocol. The case of constant velocity pulling and oscillatory driving are considered in Sec.~\ref{sec:work_calc}, and expressions for the work dissipated in these two protocols are derived. A comparison with previous work done in the field of dissipation calculation for driven Rouse chains is provided in Sec.~\ref{sec:fd_compare_approach}, and Sec.~\ref{sec:res_disc} discusses the results of the present work for both the driving protocols. We conclude in Sec.~\ref{sec:otl_concl}. The appendix contains the derivation of identities used in the simplification of expressions appearing in Sections~\ref{sec:gov_eq} and ~\ref{sec:work_calc}.

\section{\label{sec:gov_eq} Model Description and Governing equations}

A polymer chain with $(N+1)$ beads is considered in the present work, with each bead of radius $a$ connected to its nearest neighbor via a spring-dashpot arrangement as shown in Fig.~\ref{fig:bsp_chain_schem}. The polymer is suspended in a newtonian solvent of viscosity $\eta_{\text{s}}$ at a temperature $T$ and is confined to motion along a single dimension. The stiffness of each Hookean spring is $H$, and the damping coefficient of each dashpot is $K$. The positions of the beads are denoted by $\left[y_{0},y_{1},\cdots,y_{N}\right]$. The first bead of the chain is kept fixed at the origin, i.e., $y_0=0$. The last bead is attached to an optical/laser trap of stiffness $cH$, where $c$ is a positive constant, and the location of the trap $Y(t)$ is varied in time according to some protocol. The Hamiltonian of the system is therefore given by
\begin{align}
\mathcal{H}=\dfrac{H}{2}\sum_{\nu=0}^{N-1}\left(y_{\nu+1}-y_{\nu}\right)^2+\dfrac{cH}{2}\left(Y(t)-y_{N}\right)^2
\end{align}
Using the principles of polymer kinetic theory~\cite{ravibook,Bird1987b}, and assuming the beads are massless, a force-balance may be written for the $\nu^{\text{th}}$ bead in the chain, by requiring that the sum of the drag force, spring force, internal friction (IV) force, and the Brownian force vanishes. The drag force, spring force and the Brownian force on the $\nu^{\text{th}}$ bead are given by $F^{\text{D}}_{\nu}=-\zeta\llbracket\dot{y}_{\nu}\rrbracket$, $F^{\text{S}}_{\nu}=-\left(\partial \mathcal{H}/\partial y_{\nu}\right)$ and $F^{\text{B}}_{\nu}=-k_BT\left(\partial \ln \Psi/\partial y_{\nu}\right)$, respectively, where, $k_B$ is Boltzmann's constant, $\zeta=6\pi\eta_{s}a$ the drag coefficient, $\llbracket\cdots\rrbracket$ denotes a momentum-average, $\llbracket\dot{y}_{\nu}\rrbracket$ the momentum-averaged velocity of the $\nu^{\text{th}}$ bead, and $\Psi\equiv\Psi\left(y_1,\cdots y_N\right)$ is the configurational distribution function of the bead positions. The IV force on the $\nu^{\text{th}}$ that is not at the chain ends is written as follows~\cite{ravibook}
\begin{align}
F^{\text{IV}}_{\nu}=K\llbracket \dot{y}_{\nu+1}-2\dot{y}_{\nu}+\dot{y}_{\nu-1}\rrbracket
\end{align}
The equation of motion for the $\nu^{\text{th}}$ bead that is not at the chain ends may therefore be given by the following expression
\begin{align}
\llbracket\dot{y}_{\nu}\rrbracket=-\dfrac{k_BT}{\zeta}\left(\dfrac{\partial \Psi}{\partial y_{\nu}}\right)-\dfrac{1}{\zeta}\left(\dfrac{\partial \mathcal{H}}{\partial y_{\nu}}\right)+\dfrac{1}{\zeta}F^{\text{IV}}_{\nu}
\end{align}
It is convenient to work with collective coordinates $\bm{y}\equiv\left[y_1,y_2,\cdots y_{N}\right]$, so that the equation of motion may be rewritten as follows
\begin{equation}\label{eq:coll_coord_eom}
\llbracket\dot{\bm{y}}\rrbracket=-\dfrac{H}{\zeta}\bm{A}\cdot\bm{y}-\dfrac{K}{\zeta}\bm{L}\cdot\llbracket\dot{\bm{y}}\rrbracket+\dfrac{cH}{\zeta}\boldsymbol{\chi}(t)-\dfrac{k_BT}{\zeta}\left(\dfrac{\partial \ln\Psi}{\partial \bm{y}}\right)
\end{equation}
where $\boldsymbol{\chi}(t)=[0,0,\dots,Y(t)]$ represents the deterministic pulling-protocol applied to the steering bead; $\bm{A}$ and $\bm{L}$ are symmetric tridiagonal $N\times N$ matrices whose elements are defined as follows
 \begin{align}\label{eq:a_matrix_def}
A_{jk}= \left\{
\begin{array}{ll}
       2; &  j=k\neq N \\[5pt]
       (c+1); &  j=k=N \\[5pt]
      -1; & |j-k|=1 \\[5pt]
       0 ; & \text{otherwise}
\end{array} 
\right. 
\end{align}
 \begin{align}\label{eq:l_matrix_def}
L_{jk}= \left\{
\begin{array}{ll}
       2; &  j=k\neq N \\[5pt]
       1; &  j=k=N \\[5pt]
      -1; & |j-k|=1 \\[5pt]
       0 ; & \text{otherwise}
\end{array} 
\right. 
\end{align}
Note that $\bm{A}$ becomes the standard Rouse matrix when the stiffness of the steering spring is the same as that of the springs in the polymer, i.e., for $c=1$.
Upon simplifying eq.~(\ref{eq:coll_coord_eom}),
\begin{equation}\label{eq:decoup_eom}
\llbracket\dot{\bm{y}}\rrbracket=-\dfrac{H}{\zeta}\bm{D}\cdot\bm{A}\cdot\bm{y}+\dfrac{cH}{\zeta}\bm{D}\cdot\boldsymbol{\chi}(t)-\dfrac{k_BT}{\zeta}\bm{D}\cdot\left(\dfrac{\partial \ln\Psi}{\partial \bm{y}}\right),
\end{equation} 
with
\begin{equation}\label{eq:d_mat_def}
\bm{D}=\left[\boldsymbol{\delta}+\varphi\bm{L}\right]^{-1},
\end{equation}
where $\varphi=K/\zeta$ is the internal friction parameter. Substituting eq.~(\ref{eq:decoup_eom}) into the equation of continuity for the probability density in configuration space,
\begin{equation}
\dfrac{\partial \Psi}{\partial t}=-\dfrac{\partial}{\partial \bm{y}}\cdot\left\{\llbracket\dot{\bm{y}}\rrbracket\Psi\right\}
\end{equation}
we obtain the following Fokker-Planck equation
\begin{align}\label{eq:fp_ddot}
\dfrac{\partial \Psi}{\partial t}=&-\dfrac{\partial}{\partial \bm{y}}\cdot\left\{\left[-\dfrac{H}{\zeta}\bm{D}\cdot\bm{A}\cdot\bm{y}+\dfrac{cH}{\zeta}\bm{D}\cdot\boldsymbol{\chi}(t)\right]\Psi\right\}\\\nonumber
&+\dfrac{1}{2}\left(\dfrac{2k_BT}{\zeta}\right)\dfrac{\partial}{\partial \bm{y}}\dfrac{\partial}{\partial \bm{y}}:\left[\bm{D}\Psi\right],
\end{align}
where we have made use of the fact that $\left(\partial/\partial \bm{y}\right)\cdot\bm{D}=0$ since the diffusion tensor $\bm{D}$ for the system is independent of bead positions.
The stochastic differential equation equivalent to eq.~(\ref{eq:fp_ddot}) may be written using the It\^{o} interpretation~\cite{Ottinger1996} as follows
\begin{align}\label{eq:gov_sde}
d\bm{y}&=\left[-\dfrac{H}{\zeta}\bm{D}\cdot\bm{A}\cdot\bm{y}+\dfrac{cH}{\zeta}\bm{D}\cdot\boldsymbol{\chi}(t)\right]dt\nonumber\\
&+\sqrt{\dfrac{2k_BT}{\zeta}}\bm{B}\cdot d\bm{M}_{t};\quad\bm{B}\cdot\bm{B}^{T}=\bm{D}
\end{align}
where $\bm{M}_t$ represents a Wiener process.
The equivalent Langevin form of eq.~(\ref{eq:gov_sde}) is given by
\begin{equation}\label{eq:langevin_gov}
\dfrac{d\bm{y}}{dt}=\left(\frac{H}{\zeta}\right)\left[-\bm{D}\cdot\bm{A}\cdot\bm{y}+c\bm{D}\cdot\boldsymbol{\chi}(t)\right]+\sqrt\dfrac{{k_BT}}{{\zeta}}\boldsymbol{\eta}(t)
\end{equation}
where the moments of the noise term $\boldsymbol{\eta}(t)$ obey
\begin{equation}\label{eq:dim_noise}
\left<\boldsymbol{\eta}(t)\right>=0;\quad \left<\boldsymbol{\eta}(t')\boldsymbol{\eta}(t'')\right>=2\bm{D}\delta(t'-t'')
\end{equation}

In our subsequent calculations, we use the length scale $l_{H}=\sqrt{k_BT/H}$ and time scale $t_{\text{s}}=\zeta/H$ to non-dimensionalize eq.~(\ref{eq:langevin_gov}) and obtain
\begin{align}\label{eq:dimless_langevin}
\dfrac{d\bm{z}}{d\tau}=-\bm{D}\cdot\bm{A}\cdot\bm{z}(\tau)+\bm{D}\cdot\boldsymbol{h}(\tau)+\boldsymbol{\mu}(\tau)
\end{align}
where
\begin{equation}\label{eq:def_prop_1}
\begin{split}
\bm{z}&={\bm{y}}/{l_{H}}=\sqrt{{H}/{k_BT}}\,\bm{y};\quad
\alpha={Y}/{l_{H}}=\sqrt{{H}/{k_BT}}\,Y;\\[10pt]
\tau&={Ht}/{\zeta}; \quad
\bm{h}(\tau)=[0,0,\cdots,c\alpha(\tau)]
\end{split}
\end{equation}
and the dimensionless noise term obeys
\begin{equation}\label{eq:mean_noise}
\left<\boldsymbol{\mu}(\tau)\right>=0;
\end{equation}
and
\begin{equation}\label{eq:var_noise}
\left<\boldsymbol{\mu}(\tau')\boldsymbol{\mu}(\tau'')\right>=2\bm{D}\delta(\tau'-\tau'').
\end{equation}
Except for the trivial case of $\varphi=0$, the matrix $\bm{D.A}$ is not symmetric even though $\bm{D}$ and $\bm{A}$ are symmetric, since the two matrices do not commute in general, i.e., $\bm{D}\cdot\bm{A}\neq\bm{A}\cdot\bm{D}$. When $\varphi=0$, $\bm{D}$ is simply the identity matrix and the Rouse chain result~\cite{Dhar2005} is recovered, as discussed later in Sec.~\ref{sec:fd_compare_approach}.
The formal solution to eq.~(\ref{eq:dimless_langevin}) is given by
\begin{align}\label{eq:formal_z}
&\bm{z}(\tau)=\bm{G}(\tau)\cdot\bm{z}(0)\\\nonumber
&+\int_{0}^{\tau}d\tau'\bm{G}(\tau-\tau')\cdot\left[\bm{D}\cdot\boldsymbol{h}(\tau')+\boldsymbol{\mu}(\tau')\right]
\end{align}
where
\begin{equation}\label{eq:g_mat_def}
\bm{G}(\tau)=\exp\left[-\bm{D}\cdot\bm{A}\tau\right]
\end{equation}
The Jarzynski work~\cite{Jarzynski1997,Dhar2005} done over a time interval $\tau_{\text{m}}$ may be evaluated as follows,
\begin{align}
&{W}=\dfrac{W^{*}}{k_BT}=\dfrac{1}{k_BT}\int_{0}^{t_{\text{m}}}\dfrac{\partial \mathcal{H}}{\partial Y}\dot{Y}dt=c\int_{0}^{\tau_{\text{m}}}\left(\alpha-z_{N}\right)\dot{\alpha}d\tau,
\end{align}
giving
\begin{equation}\label{eq:w_j_rel}
{W}=\dfrac{c}{2}\left[\alpha^2(\tau_{\text{m}})-\alpha^2(0)\right]-\int_{0}^{\tau_{\text{m}}}d\tau\dot{\bm{h}}(\tau)\cdot\bm{z}(\tau)
\end{equation}
Substituting eq.~(\ref{eq:formal_z}) into eq.~(\ref{eq:w_j_rel}), we obtain
\begin{align}\label{eq:work_rel_exp}
&{W}=\dfrac{c}{2}\left[\alpha^2(\tau_{\text{m}})-\alpha^2(0)\right]-\int_{0}^{\tau_{\text{m}}}d\tau\dot{\bm{h}}(\tau)\cdot\Biggl[\bm{G}(\tau)\cdot\bm{z}(0)\nonumber\\[5pt]
&+\int_{0}^{\tau}d\tau'\bm{G}(\tau-\tau')\cdot\bm{D}\cdot\boldsymbol{h}(\tau')+\int_{0}^{\tau}d\tau'\bm{G}(\tau-\tau')\cdot\boldsymbol{\mu}(\tau')\Biggr]
\end{align}
A matrix representation of the Hamiltonian may be written as follows
\begin{align}\label{eq:hamilt_def}
\dfrac{\mathcal{H}}{k_BT}=\dfrac{1}{2}\bm{z}\cdot\bm{A}\cdot\bm{z}+\dfrac{c}{2}\alpha^2-\bm{h}\cdot\bm{z}
\end{align}
Using eq.~(\ref{eq:hamilt_def}), the dimensionless partition function is evaluated as
\begin{align}\label{eq:z_part_def}
{\mathcal{Z}}&=\int\exp\left[-\dfrac{\mathcal{H}}{k_BT}\right]d\bm{z}\\\nonumber
&=\exp\left[-\dfrac{c\alpha^2}{2}\right]\int\exp\left[-\dfrac{1}{2}\bm{A}:\bm{z}\bm{z}+\bm{h}\cdot\bm{z}\right]d\bm{z}
\end{align}
The integral in eq.~(\ref{eq:z_part_def}) may be simplified using the identity E.3-7 in ref.~\citenum{Bird1987b} to obtain the following expression for the dimensionless partition function,
\begin{equation}\label{eq:part_func_final}
{\mathcal{Z}}=\dfrac{(2\pi)^{N/2}}{\sqrt{\det(\bm{A})}}\exp\left[-\dfrac{c\alpha^2}{2}+\dfrac{1}{2}\bm{h}\cdot\bm{A}^{-1}\cdot\bm{h}\right]
\end{equation}
The dimensionless free energy is then obtained as follows
\begin{equation}\label{eq:free_energy}
{F}[\alpha(\tau)]=-\ln{\mathcal{Z}}=\dfrac{c\alpha^2(\tau)}{2}-\dfrac{1}{2}\bm{h}(\tau)\cdot\bm{A}^{-1}\cdot\bm{h}(\tau)
\end{equation}
after ignoring the constant terms.
The dimensionless probability distribution function for the bead positions in the system is given by
\begin{align}
&\Psi(\bm{z})=\dfrac{1}{{\mathcal{Z}}}\exp\left[-\dfrac{\mathcal{H}}{k_BT}\right]\\\nonumber
&=\dfrac{\sqrt{\det(\bm{A})}}{(2\pi)^{N/2}}\exp\left\{-\dfrac{1}{2}\bm{z}\cdot\bm{A}\cdot\bm{z}-\dfrac{1}{2}\bm{h}\cdot\bm{A}^{-1}\cdot\bm{h}+\bm{h}\cdot\bm{z}\right\}
\end{align}
The mean positions of the beads and their fluctuations may then be deduced as follows
\begin{equation}\label{eq:mean_pos}
\left<\bm{z}\right>=\bm{A}^{-1}\cdot\bm{h}
\end{equation}
\begin{equation}\label{eq:var_pos}
\left<\left(\bm{z}-\left<\bm{z}\right>\right)\left(\bm{z}-\left<\bm{z}\right>\right)\right>=\bm{A}^{-1}
\end{equation}
Taking an ensemble average on both sides of eq.~(\ref{eq:work_rel_exp}), the expression for the mean work is obtained as follows
\begin{align}\label{eq:mean_work_step2}
&\left<{W}\right>=\dfrac{c}{2}\left[\alpha^2(\tau_{\text{m}})-\alpha^2(0)\right]\nonumber\\
&-\int_{0}^{\tau_{\text{m}}}d\tau\dot{\bm{h}}(\tau)\cdot\bm{G}(\tau)\cdot\bm{A}^{-1}\cdot\bm{h}(0)\nonumber\\[5pt]
&-\int_{0}^{\tau_{\text{m}}}d\tau\dot{\bm{h}}(\tau)\cdot\int_{0}^{\tau}d\tau'\cdot\bm{G}(\tau-\tau')\cdot\bm{D}\cdot\boldsymbol{h}(\tau')
\end{align}
where we have used eq.~(\ref{eq:mean_pos}) and the fact that the ensemble-average of the dimensionless noise term vanishes (cf. eq.~(\ref{eq:mean_noise})). The last term on the RHS of eq.~(\ref{eq:mean_work_step2}) is simplified using integration by parts as shown in eqs.~(\ref{eq:def_i}) - (\ref{eq:i_final}) of the Appendix to obtain
\begin{align}\label{eq:av_work_intermed}
\left<{W}\right>&=\dfrac{c}{2}\left[\alpha^2(\tau_{\text{m}})-\alpha^2(0)\right]-\int_{0}^{\tau_{\text{m}}}d\tau\dot{\bm{h}}(\tau)\cdot\bm{A}^{-1}\cdot\bm{h}(\tau)\nonumber\\[5pt]
&+\int_{0}^{\tau_{\text{m}}}d\tau\int_{0}^{\tau}d\tau'\dot{\bm{h}}(\tau)\cdot\bm{G}(\tau-\tau')\cdot\bm{A}^{-1}\cdot\dot{\bm{h}}(\tau')
\end{align}
Simplifying the second term on the RHS of eq.~(\ref{eq:av_work_intermed}) using integration by parts as shown in eqs.~(\ref{eq:def_j}) - ~(\ref{eq:j_ans}) of the Appendix results in
\begin{align}\label{eq:w_av_simp2}
\left<{W}\right>&=\left[\dfrac{c}{2}\alpha^2(\tau_{\text{m}})-\dfrac{1}{2}\bm{h}(\tau_{\text{m}})\cdot\bm{A}^{-1}\cdot\bm{h}(\tau_{\text{m}})\right]\nonumber\\[5pt]
&-\left[\dfrac{c}{2}\alpha^2(0)-\dfrac{1}{2}\bm{h}(0)\cdot\bm{A}^{-1}\cdot\bm{h}(0)\right]\nonumber\\[5pt]
&+\int_{0}^{\tau_{\text{m}}}d\tau\int_{0}^{\tau}d\tau'\dot{\bm{h}}(\tau)\cdot\bm{G}(\tau-\tau')\cdot\bm{A}^{-1}\cdot\dot{\bm{h}}(\tau').
\end{align}
Recognizing that the first two terms on the RHS of the above equation represent the free-energy difference for the transition between the initial and final states (cf. eq.~(\ref{eq:free_energy})), we may write
\begin{align}\label{eq:w_components}
\left<{W}\right>&=\Delta{F}+\left<{W}_{\text{dis}}\right>
\end{align}
where
\begin{equation}\label{eq:delta_f_relation}
\Delta{F}={F}\left[\alpha(\tau_{\text{m}})\right]-{F}\left[\alpha(0)\right],
\end{equation}
and
\begin{equation}\label{eq:wdis_relation}
\left<{W}_{\text{dis}}\right>=\int_{0}^{\tau_{\text{m}}}d\tau\int_{0}^{\tau}d\tau'\dot{\bm{h}}(\tau)\cdot\bm{G}(\tau-\tau')\cdot\bm{A}^{-1}\cdot\dot{\bm{h}}(\tau')
\end{equation}
We have therefore identified, in eq.~(\ref{eq:w_components}), the reversible (free-energy) and irreversible (dissipative) components of the total average work done during the transition. {By definition, the free-energy difference between the terminal states is the work done in quasi-statically driving the system between the two states~\cite{Callen1985}.} We proceed to calculate the variance associated with the work process, defined as
\begin{equation}\label{eq:var_def}
\sigma^2=\left<\left(W-\left<{W}\right>\right)^2\right>
\end{equation}
Subtracting eq.~(\ref{eq:mean_work_step2}) from eq.~(\ref{eq:work_rel_exp}) and simplifying, we obtain
\begin{widetext}
\begin{equation}\label{eq:diff_work}
\begin{split}
W-\left<{W}\right>=-\int_{0}^{\tau_{\text{m}}}d\tau\dot{\bm{h}}(\tau)\cdot\bm{G}(\tau)\cdot\left[\bm{z}(0)-\left<\bm{z}(0)\right>\right]-\int_{0}^{\tau_{\text{m}}}d\tau\int_{0}^{\tau}d\tau'\dot{\bm{h}}(\tau)\cdot\bm{G}(\tau-\tau')\cdot\boldsymbol{\mu}(\tau')
\end{split}
\end{equation}
We may therefore use eq.~(\ref{eq:diff_work}) to calculate $\sigma^2$ as follows
\begin{equation}\label{eq:sigma_intermed_1}
\begin{split}
\sigma^2&=\int_{0}^{\tau_{\text{m}}}d\tau_{1}\int_{0}^{\tau_{\text{m}}}d\tau_{2}\dot{\bm{h}}(\tau_1)\cdot\bm{G}(\tau_1)\cdot\uline{\left<\left(\bm{z}(0)-\left<\bm{z}(0)\right>\right)\left(\bm{z}(0)-\left<\bm{z}(0)\right>\right)\right>}\cdot\bm{G}(\tau_2)\cdot\dot{\bm{h}}(\tau_2)\\[10pt]
&+\int_{0}^{\tau_{\text{m}}}d\tau_{1}\int_{0}^{\tau_1}d\tau'_{1}\int_{0}^{\tau_{\text{m}}}d\tau_{2}\int_{0}^{\tau_2}d\tau'_{2}\,\dot{\bm{h}}(\tau_1)\cdot\bm{G}(\tau_1-\tau'_1)\cdot\uline{\left<\boldsymbol{\mu}(\tau'_{1})\boldsymbol{\mu}(\tau'_{2})\right>}\cdot\bm{G}(\tau_2-\tau'_2)\cdot\dot{\bm{h}}(\tau_2)
\end{split}
\end{equation}
Using eqs.~(\ref{eq:var_pos}) and ~(\ref{eq:var_noise}) to simplify the underlined terms in the first and second lines of Eq.~(\ref{eq:sigma_intermed_1}), respectively, and performing the integral over $\tau'_{1}$ on the RHS of eq.~(\ref{eq:sigma_intermed_1}), we obtain
\begin{equation}\label{eq:sigma_intermed_3}
\begin{split}
\sigma^2&=\int_{0}^{\tau_{\text{m}}}d\tau_{1}\int_{0}^{\tau_{\text{m}}}d\tau_{2}\,\dot{\bm{h}}(\tau_1)\cdot\bm{G}(\tau_1)\cdot\bm{A}^{-1}\cdot\bm{G}(\tau_2)\cdot\dot{\bm{h}}(\tau_2)\\[10pt]
&+2\int_{0}^{\tau_{\text{m}}}d\tau_{1}\int_{0}^{\tau_{\text{m}}}d\tau_{2}{\int_{0}^{\tau_2}d\tau'_{2}\,\dot{\bm{h}}(\tau_1)\cdot\bm{G}(\tau_1-\tau'_2)\cdot\bm{D}\cdot\bm{G}(\tau_2-\tau'_2)\cdot\dot{\bm{h}}(\tau_2)}
\end{split}
\end{equation}
Using the following identity derived in eqs.~(\ref{eq:aside_int1})-(\ref{eq:aside_int2}) of the Appendix,
\begin{align}\label{eq:aside_int3}
\int_{0}^{\tau_2}d\tau'_{2}\,\bm{G}(\tau_1-\tau'_2)\cdot\bm{D}\cdot\bm{G}(\tau_2-\tau'_2)=\dfrac{1}{2}\bm{A}^{-1}\cdot\bm{G}(\tau_1-\tau_2)-\dfrac{1}{2}\bm{G}(\tau_1)\cdot\bm{A}^{-1}\cdot\bm{G}(\tau_2)
\end{align}
to simplify the second term on the RHS of eq.~(\ref{eq:sigma_intermed_3}), we obtain
\begin{equation}\label{eq:variance_relation}
\sigma^2=\int_{0}^{\tau_{\text{m}}}d\tau_{1}\int_{0}^{\tau_{\text{m}}}d\tau_{2}\,\dot{\bm{h}}(\tau_1)\cdot\bm{A}^{-1}\cdot\bm{G}(\tau_1-\tau_2)\cdot\dot{\bm{h}}(\tau_2)=2\int_{0}^{\tau_{\text{m}}}d\tau_{1}\int_{0}^{\tau_{1}}d\tau_{2}\,\dot{\bm{h}}(\tau_1)\cdot\bm{A}^{-1}\cdot\bm{G}(\tau_1-\tau_2)\cdot\dot{\bm{h}}(\tau_2)
\end{equation}
\end{widetext}
From eqs.~(\ref{eq:wdis_relation}) and ~(\ref{eq:variance_relation}), we have
\begin{equation}\label{eq:var_wdis_rel}
\sigma^2=2\left<{W}_{\text{dis}}\right>
\end{equation}
which is expected for a Gaussian work distribution~\cite{Dhar2005,Jarzynski1997}. 

{Jarzynski~\cite{Jarzynski1997}, in 1997, derived the following equality
\begin{equation}\label{eq:jeq_ori}
\left<\exp\left[-W\right]\right>=\exp\left[-\Delta F\right]
\end{equation}
that allows for the estimation of free-energy differences from finite-rate, rather than quasistatic driving. For microscopic systems in which Brownian fluctuations are significant, a distribution of work values are expected whilst the system is being driven, and consequently, a large enough ensemble of work trajectories need to be collected to get reliable estimates. The experimental validity of Jarzynski's equality has been established for colloidal systems~\cite{Toyabe2010} and polymers~\cite{Liphardt2002} alike. The use of Hookean springs results in the work distribution being Gaussian and Jarzynski's equality [eq.~(\ref{eq:jeq_ori})] is trivially satisfied for the system and protocols discussed in the present work~\cite{Kailasham2020}. In its original form, the Jarzynski equality is valid for transitions between equilibrium steady states (ESS). A stationary colloidal particle in a harmonic trapping potential being moved from an initial to a final position, or a polymer chain whose terminii are subjected to and manipulated by optical traps are examples of systems driven between ESS. On the contrary, trapped colloidal beads moving at a fixed velocity and driven to a different velocity~\cite{Trepagnier2004}, polymer chains being driven between two values of the flow strength~\cite{Latinwo2013,Latinwo2014,Latinwo2014a,Zhou2022,Ghosal2016,Sharma2011,Ghosal2016a}, represent transitions between non-equilibrium steady states (NESS). For these cases, the Jarzynski equality has to be suitably modified~\cite{Hatano2001,Harada2005}.} 

\section{\label{sec:work_calc} Work calculation}

In this section, we calculate the work dissipated in driving the chain under two protocols that are commonly used in single-molecule force experiments~\cite{Woodside2014,Liphardt2002,Braun2004a,Szymczak2009,Wu2018}. One manoeuvre involves moving the trap position at a constant velocity over a fixed distance. We refer to this interchangeably as the constant velocity pulling or the linear protocol. The other method for driving the chain involves changing the trap position in a periodic manner. This is referred to as either the oscillatory driving or the symmetric protocol.  

\subsection{\label{sec:const_vel} Constant velocity pulling (linear protocol)}
We now evaluate the work statistics associated with a constant velocity pulling process, given by 
\begin{equation}\label{eq:const_vel_prot}
\alpha(\tau)=\alpha(0)+v\tau
\end{equation}
that acts over the interval $[0,\tau_{\text{m}}]$, such that the final bead is pulled over a dimensionless distance $[\alpha(\tau_{\text{m}})-\alpha(0)] \equiv d=v\tau_{\text{m}}$ in the process. From the definition of $\bm{h}(\tau)$ in eq.~(\ref{eq:def_prop_1}) and the protocol defined in eq.~(\ref{eq:const_vel_prot}), we may write
\begin{equation}\label{eq:simp_hdef}
\dot{\bm{h}}(\tau)=\left[0,0,\cdots,c\,v\right]
\end{equation}
The vector in eq.~(\ref{eq:simp_hdef}) is of length $N$. Substituting this value into the expression for dissipated work given by eq.~(\ref{eq:wdis_relation}), 
\begin{align}\label{eq:wexp2}
\left<{W}_{\text{dis}}\right>&=\int_{0}^{\tau_{\text{m}}}d\tau\int_{0}^{\tau}d\tau'\dot{\bm{h}}(\tau)\cdot\bm{G}(\tau-\tau')\cdot\bm{A}^{-1}\cdot\dot{\bm{h}}(\tau')\nonumber\\
&=c^2v^2\int_{0}^{\tau_{\text{m}}}d\tau\int_{0}^{\tau}d\tau'\left[\bm{G}(\tau-\tau')\cdot\bm{A}^{-1}\right]_{NN},
\end{align}
where $[\dots]_{NN}$ denotes a matrix element. Processing the integrals on the RHS of eq.~(\ref{eq:wexp2}) as indicated in eqs.~(\ref{eq:g_i_1})-~(\ref{eq:g_i_3}) of the Appendix, one obtains
\begin{widetext}
\begin{equation}\label{eq:wdis_const_vel}
\begin{split}
&\left<{W}_{\text{dis}}\right>=\dfrac{c^2d^2}{\tau_{\text{m}}}\Biggl\{\bm{A}^{-1}\cdot\bm{D}^{-1}\cdot\bm{A}^{-1}+\dfrac{1}{\tau_{\text{m}}}\left(\bm{A}^{-1}\cdot\bm{D}^{-1}\right)\cdot\left[\exp\left[-\bm{D}\cdot\bm{A}\tau_{\text{m}}\right]-\boldsymbol{\delta}\right]\cdot\bm{A}^{-1}\cdot\left(\bm{D}^{-1}\cdot\bm{A}^{-1}\right)\Biggr\}_{NN}
\end{split}
\end{equation}
\subsection{\label{oscill_drive} Oscillatory driving (symmetric protocol)}
We next calculate the work done in oscillatory driving, i.e., assuming that the trap location is moved according to 
\begin{equation}
\alpha(\tau)=\alpha(0)+d\sin(\omega\tau)
\end{equation}
such that the trap location moves a dimensionless distance of $d$ in a dimensionless time interval of $\tau_{\text{m}}$. This requires
\begin{equation}\label{eq:wt_cons}
\alpha(\tau_{\text{m}})-\alpha(0)=d\sin(\omega\tau_{\text{m}})=d\implies \omega=\pi/(2\tau_{\text{m}}).
\end{equation}
We therefore have
\begin{equation}\label{eq:oscill_hdef}
\bm{h}(\tau)=\left[0,0,\cdots,c\,\alpha(\tau)\right];\,\dot{\bm{h}}(\tau)=\left[0,0,\cdots,cd\omega\cos(\omega\tau)\right]
\end{equation}
Substituting the above pulling protocol into the expression for dissipated work given by eq.~(\ref{eq:wdis_relation}), we obtain an expression for the averaged dissipated work as follows
\begin{align}\label{eq:wdis_int_oscil}
&\left<{W}_{\text{dis}}\right>=\int_{0}^{\tau_{\text{m}}}d\tau\int_{0}^{\tau}d\tau'\dot{\bm{h}}(\tau)\cdot\bm{G}(\tau-\tau')\cdot\bm{A}^{-1}\cdot\dot{\bm{h}}(\tau')=c^2d^2\omega^2\int_{0}^{\tau_{\text{m}}}d\tau\cos(\omega\tau)\int_{0}^{\tau}d\tau'\left[\cos(\omega\tau')\bm{G}(\tau-\tau')\cdot\bm{A}^{-1}\right]_{NN}
\end{align}
Simplifying the RHS of Eq.~(\ref{eq:wdis_int_oscil}) requires several rounds of integration by parts, as illustrated in eqs.~(\ref{eq:os_1})-~(\ref{eq:outer_intg_oscil}) of the Appendix, resulting in the following expression
\begin{equation}\label{eq:wdis_o_fin}
\begin{split}
\left<W_{\text{dis}}\right>&=\dfrac{c^2d^2\omega^2}{4}\left[1-\cos(2\omega\tau_{\text{m}})\right]\left[\boldsymbol{\Phi}\cdot\bm{A}^{-1}\right]_{NN}+\dfrac{c^2d^2\omega}{4}\left[2\omega\tau_{\text{m}}+\sin(2\omega\tau_{\text{m}})\right]\left[\boldsymbol{\Phi}\cdot\bm{D}\right]_{NN}\\[5pt]
&-c^2d^2\omega^3\left[\sin(\omega\tau_{\text{m}})\boldsymbol{\Phi}^2\cdot\bm{G}(\tau_{\text{m}})\cdot\bm{D}\right]_{NN}+c^2d^2\omega^2\left[\cos(\omega\tau_{\text{m}})\boldsymbol{\Phi}^2\cdot\left(\bm{D}\cdot\bm{A}\right)\cdot\bm{G}(\tau_{\text{m}})\cdot\bm{D}\right]_{NN}\\[5pt]
&-c^2d^2\omega^2\left[\boldsymbol{\Phi}^2\cdot\bm{D}\cdot\bm{A}\cdot\bm{D}\right]_{NN}
\end{split}
\end{equation}
where $\boldsymbol{\Phi}=\left[\omega^2\boldsymbol{\delta}+\left(\bm{D}\cdot\bm{A}\right)^2\right]^{-1}$.
We have required for the current protocol that $\omega\tau_{\text{m}}=\pi/2$, as illustrated in eq.~(\ref{eq:wt_cons}). With this simplification, eq.~(\ref{eq:wdis_o_fin}) becomes
\begin{align}\label{eq:wdis_o_num}
\left<W_{\text{dis}}\right>=\dfrac{c^2d^2\omega^2}{2}\left[\boldsymbol{\Phi}\cdot\bm{A}^{-1}\right]_{NN}+\dfrac{c^2d^2\omega\pi}{4}\left[\boldsymbol{\Phi}\cdot\bm{D}\right]_{NN}-c^2d^2\omega^3\left[\boldsymbol{\Phi}^2\cdot\bm{G}(\tau_{\text{m}})\cdot\bm{D}\right]_{NN}-c^2d^2\omega^2\left[\boldsymbol{\Phi}^2\cdot\bm{D}\cdot\bm{A}\cdot\bm{D}\right]_{NN}
\end{align}
\end{widetext}

 \section{\label{sec:fd_compare_approach} Comparison against prior results}
 
 We now compare eq.~(\ref{eq:wdis_const_vel}) against results for two special cases which are available in the literature. Dhar~\cite{Dhar2005} derived an expression for the variance of the work fluctuations for a Rouse chain without internal friction ($K=0$) subjected to constant velocity pulling, with the steering spring of the same stiffness as the springs in the chain ($c=1$).
 
Setting $K=0$ in eq.~(\ref{eq:d_mat_def}) and $c=1$ in eq.~(\ref{eq:wdis_const_vel}), we obtain $\bm{D}=\boldsymbol{\delta}$, and
\begin{equation*}
\sigma^2_{\text{Rouse}}=\dfrac{2d^2}{\tau_{\text{m}}}\left\{\bm{A}^{-2}+\dfrac{1}{\tau_{\text{m}}}\bm{A}^{-3}\cdot\left[\exp\left[-\bm{A}\tau_{\text{m}}\right]-\boldsymbol{\delta}\right]\right\}_{NN} 
\end{equation*}
which corresponds identically to eq.~(19) in ref.~\citenum{Dhar2005}, where the author has used the notation $a$ (instead of $d$ as done in the present work)  to denote the pulling distance. 

\begin{figure*}[t]
\begin{center}
\begin{tabular}{c c}
\includegraphics[width=3.2in,height=!]{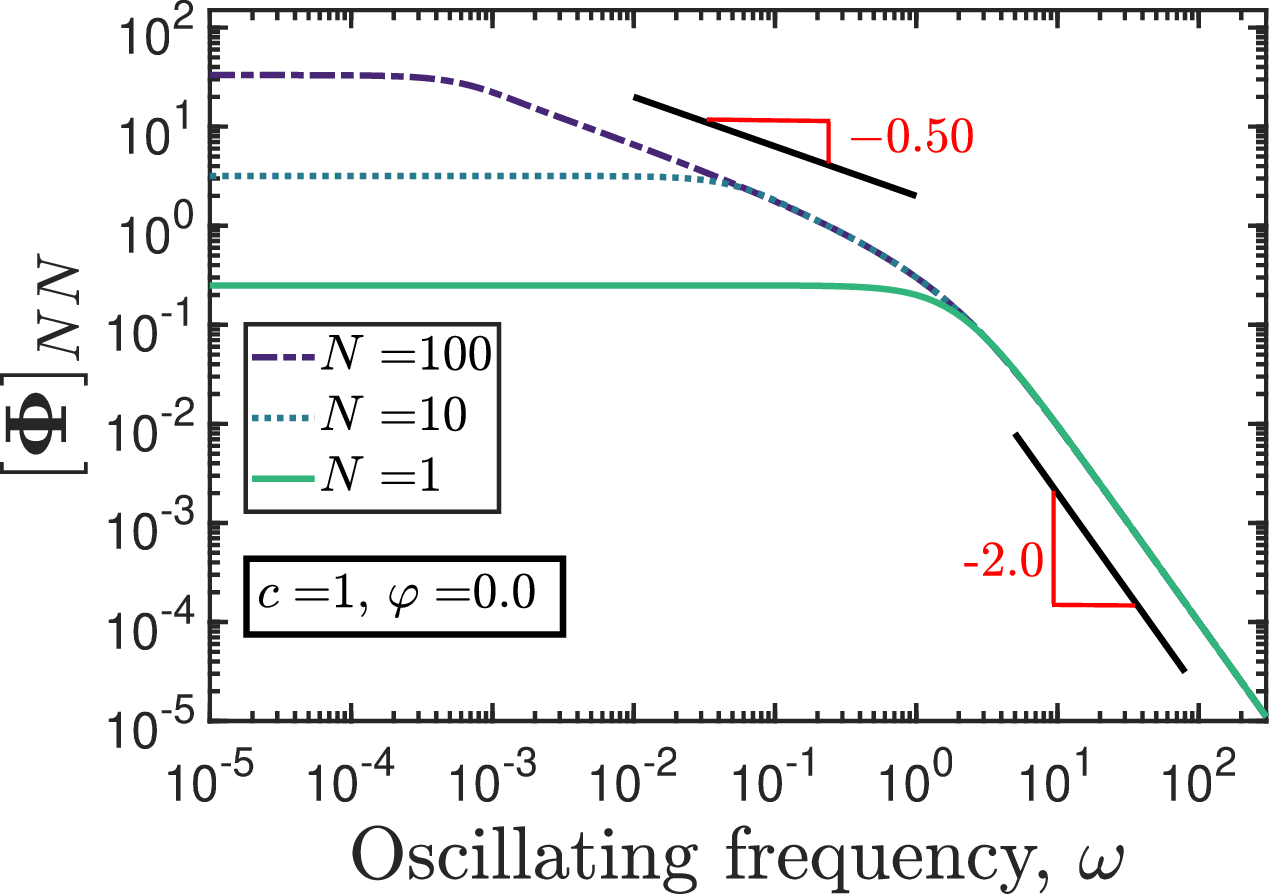}&
\includegraphics[width=3.2in,height=!]{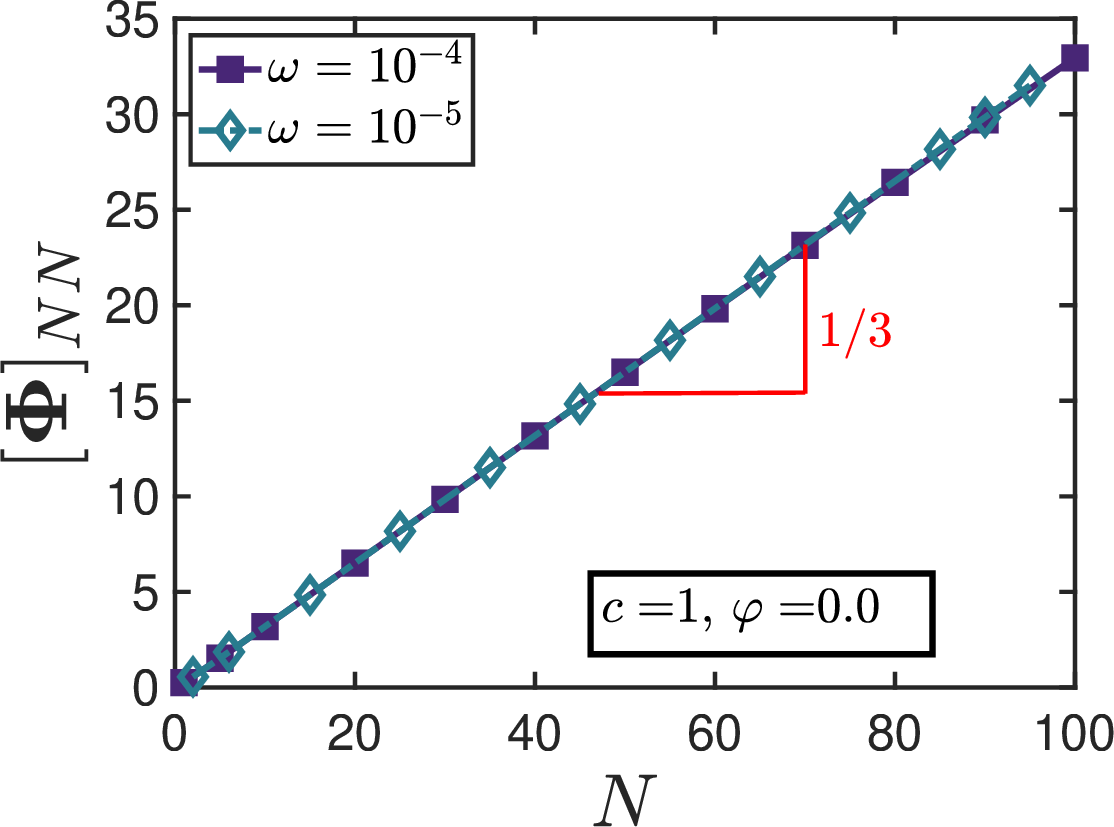}\\
(a)&(b)\\
\end{tabular}
\end{center}
\caption{The $(NN)^{\text{th}}$ term of the $\boldsymbol{\Phi}$ matrix for a Rouse chain ($\varphi=0$) pulled over a distance $d=1$ using a steering spring of stiffness $c=1$ as function of (a) driving frequency and (b) chain length. The values in (b) are evaluated for discrete values of $N$, and the line is a guide to the eye.}
\label{fig:phi_var}
\end{figure*}

The work statistics for a single-mode bead-spring-dashpot subjected to constant velocity pulling was derived in ref.~\citenum{Kailasham2020}, in which the stiffness of the steering spring was taken to be $c_2$. Setting $N=1$ and $c=c_2$, we obtain, 
\begin{equation}\label{eq:n1_sim}
\begin{split}
\left[\bm{A}\right]_{NN}&=(c_2+1)\\
\left[\bm{L}\right]_{NN}&=1\\
\left[\bm{D}\right]_{NN}&=\left[1+\left(\dfrac{K}{\zeta}\right)\right]^{-1}=\left[\dfrac{\zeta}{\zeta+K}\right]
\end{split}
\end{equation}
and the dimensionless work
\begin{align*}
&\left<{W}_{\text{dis}}\right>_{N=1}=\dfrac{\left(\zeta+K\right)}{\zeta}\left(\dfrac{c_2}{c_2+1}\right)^2vd\nonumber\\
&+\dfrac{c_2^2}{\left(c_2+1\right)^3}\left(\dfrac{\zeta+K}{\zeta}\right)^2\left[\exp\left(-\dfrac{\zeta\left(c_2+1\right)\tau_{\text{m}}}{\left(\zeta+K\right)}\right)-1\right]v^2
\end{align*}
Denoting the dimensional variables with an asterisk, we have $d=d^{*}\sqrt{H/k_BT}$, $\tau_{\text{m}}=Ht^{*}_{\text{m}}/\zeta$ and $v=v^{*}\zeta/\sqrt{k_BTH}$, we obtain
\begin{align}
&\left<{W}^{*}_{\text{dis}}\right>_{N=1}=\left(\dfrac{c_2}{c_2+1}\right)^2\left(\zeta+K\right)v^{*}d^{*}\\\nonumber
&+\dfrac{c^2_2}{\left(c_2+1\right)^3}\dfrac{\left(\zeta+K\right)^2v^{*2}}{H}\left[\exp\left(-\dfrac{H\left(c_2+1\right)d^{*}}{\left(\zeta+K\right)v^{*}}\right)-1\right],
\end{align}
which agrees identically with eq.~(16) in ref.~\citenum{Kailasham2020}. We report the value of the dimensionless dissipation for a dumbbell subjected to constant velocity pulling in the limit of large trap stiffness here: 
\begin{align}\label{eq:vel_N1}
\lim_{c\to\infty}\left<{W}_{\text{dis}}\right>_{N=1}=\left(1+\varphi\right)vd
\end{align}
as it will be useful for later discussion in the manuscript.

\vspace{-10pt}

~\citet{Speck2005} have studied Rouse chains (without internal friction) subjected to constant velocity pulling and oscillatory driving. In their paper, the last bead of the Rouse chain is directly manipulated to move at a prescribed velocity (or frequency) and is not attached to a steering spring. Their results may be extended to describe the case of a chain being pulled by a steering spring with a different stiffness, but \citet{Speck2005} only report results for the case where all the springs are of the same stiffness. To compare our results with theirs, we set $c=1$, and note for a Rouse chain that 
\begin{align}
\begin{split}
\bm{D}&=\bm{D}^{-1}=\boldsymbol{\delta}\\[5pt]
\boldsymbol{\Phi}&=\left[\omega^2\boldsymbol{\delta}+\bm{A}^{2}\right]^{-1}
\end{split}
\end{align}
resulting in the following expression for the dissipation
\begin{equation}\label{eq:wdis_rouse_oscil}
\begin{split}
&\left<W_{\text{dis}}\right>_{\text{Rouse}}=\dfrac{d^2\omega^2}{4}\left[1-\cos(2\omega\tau_{\text{m}})\right]\left[\boldsymbol{\Phi}\cdot\bm{A}^{-1}\right]_{NN}\\[5pt]
&+\dfrac{d^2\omega}{4}\left[2\omega\tau_{\text{m}}+\sin(2\omega\tau_{\text{m}})\right]\left[\boldsymbol{\Phi}\right]_{NN}\\[5pt]
&-d^2\omega^3\left[\sin(\omega\tau_{\text{m}})\boldsymbol{\Phi}^2\cdot\bm{G}(\tau_{\text{m}})\right]_{NN}\\[5pt]
&+d^2\omega^2\left[\cos(\omega\tau_{\text{m}})\boldsymbol{\Phi}^2\cdot\bm{A}\cdot\bm{G}(\tau_{\text{m}})\right]_{NN}-d^2\omega^2\left[\boldsymbol{\Phi}^2\cdot\bm{A}\right]_{NN}
\end{split}
\end{equation} 

We next seek to obtain an expression for the dissipation in the limit of slow driving $(\omega\ll1)$. In the absence of an analytical expression for $\boldsymbol{\Phi}$, we numerically evaluate the $(NN)^{\text{th}}$ element of this matrix and plot its dependence on frequency and chain length in Fig.~\ref{fig:phi_var}. 

We observe that $\boldsymbol{\Phi}_{NN}$ has a low frequency plateau whose value varies linearly with the chain length. In the slow driving regime, therefore, we may only retain terms from eq.~(\ref{eq:wdis_rouse_oscil}) that are linear in $\omega$, and reject the higher-order terms, to give
\begin{align}\label{eq:ps_slow_oscil}
&\lim_{\omega\ll1}\left<W_{\text{dis}}\right>_{\text{Rouse}}\approx\dfrac{d^2\omega}{4}\left[2\omega\tau_{\text{m}}+\sin(2\omega\tau_{\text{m}})\right]\left[\boldsymbol{\Phi}\right]_{NN}\nonumber\\[5pt]
&=\dfrac{N}{3}\dfrac{d^2\omega}{4}\left[2\omega\tau_{\text{m}}+\sin(2\omega\tau_{\text{m}})\right]
\end{align}
where the equality in the second line follows from Fig.~\ref{fig:phi_var}~(b), by replacing the $(NN)^{\text{th}}$ element of $\boldsymbol{\Phi}$ with $(N/3)$ in the limit of slow driving. 
Using $d=d^{*}\sqrt{H/k_BT}$, $\tau_{\text{m}}=Ht^{*}_{\text{m}}/\zeta$ and $\omega=\omega^{*}\zeta/H$ to cast eq.~(\ref{eq:ps_slow_oscil}) in its dimensional form, 
\begin{align}\label{eq:ps_slow_dim}
&\left<W^{*}_{\text{dis}}\right>_{\text{Rouse}}=\dfrac{N\zeta}{3}\dfrac{d^{2*}\omega^{*}}{4}\left[2\omega^*\tau^*_{\text{m}}+\sin(2\omega^*\tau^*_{\text{m}})\right]
\end{align}
We reproduce below eq.~(66) of the paper by~\citet{Speck2005}, which is an analytical expression for the average dimensional dissipation in the slow driving regime:
\begin{align}\label{eq:slow_66}
\overline{W_{\text{d}}}=\dfrac{N\gamma}{3}\dfrac{L^2\omega}{4}\left[2\omega t_{\text{s}}+\sin(2\omega t_{\text{s}})\right]
\end{align}
where $\gamma$ is the friction coefficient, which is denoted by $\zeta$ in this manuscript. The term $L$ represents the distance over which the last bead is moved over a time interval $t_{\text{s}}$, whose equivalents are $d^{*}$ and $\tau^{*}_{\text{m}}$ in the present work.
The excellent agreement between eqs.~(\ref{eq:ps_slow_dim}) and ~(\ref{eq:slow_66}) provides a validation of the work expression derived in this manuscript.

It is also instructive to write the expression for dissipation incurred in subjecting a single-mode spring-dashpot ($N=1$) to oscillatory driving, which will be compared against the dissipation incurred in driving a series of bead-spring-dashpots ($N>1$) later in the manuscript. Proceeding as indicated in eq.~(\ref{eq:n1_phi}), the following expression is obtained:
\begin{widetext}
\begin{equation}\label{eq:n1_wdis_os}
\begin{split}
&\left<W_{\text{dis}}\right>_{N=1}=\dfrac{c^2d^2\omega^2}{2}\left(\dfrac{1}{c+1}\right)\left[\dfrac{\left(1+\varphi\right)^2}{\omega^2\left(1+\varphi\right)^2+(c+1)^2}\right]+\dfrac{\pi c^2d^2\omega}{4}\left[\dfrac{\left(1+\varphi\right)}{\omega^2\left(1+\varphi\right)^2+(c+1)^2}\right]\\[5pt]
&-\dfrac{c^2d^2\omega^3\left(1+\varphi\right)^3}{\left[\omega^2\left(1+\varphi\right)^2+(c+1)^2\right]^2}\exp\left[-\dfrac{c+1}{1+\varphi}\tau_{\text{m}}\right]-\dfrac{c^2d^2\omega^2\left(1+\varphi\right)^2(c+1)}{\left[\omega^2\left(1+\varphi\right)^2+(c+1)^2\right]^2}
\end{split}
\end{equation}
In the limit of large steering spring stiffness, we note that 
\begin{align}\label{eq:oscil_large_c}
\lim_{c\to\infty}\left<W_{\text{dis}}\right>_{N=1}=\dfrac{\pi d^2\omega}{4}(1+\varphi)
\end{align} 
\end{widetext}

 \section{\label{sec:res_disc} Results and discussion} 
 
 \begin{figure}
\centering
\includegraphics[width=80mm]{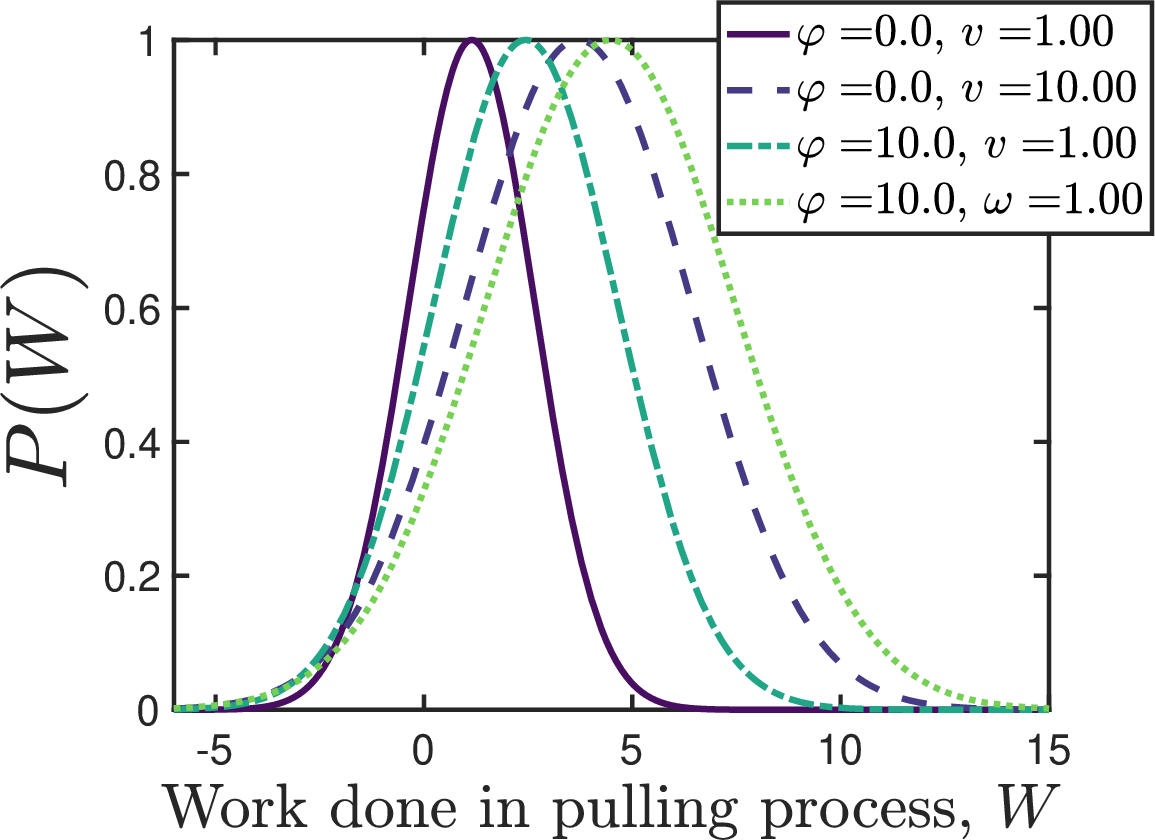}  
\caption{Probability distribution of work for a chain with $N=100$ subjected to various pulling protocols over a distance of $d=1$, using a steering spring of stiffness $c=10$. The beginning position of the trap is set to the origin for all the cases, i.e., $\alpha(0)=0$.}
\label{fig:prob_w}
\end{figure}

\begin{figure}[t]
\begin{center}
\begin{tabular}{c}
\includegraphics[width=3.in,height=!]{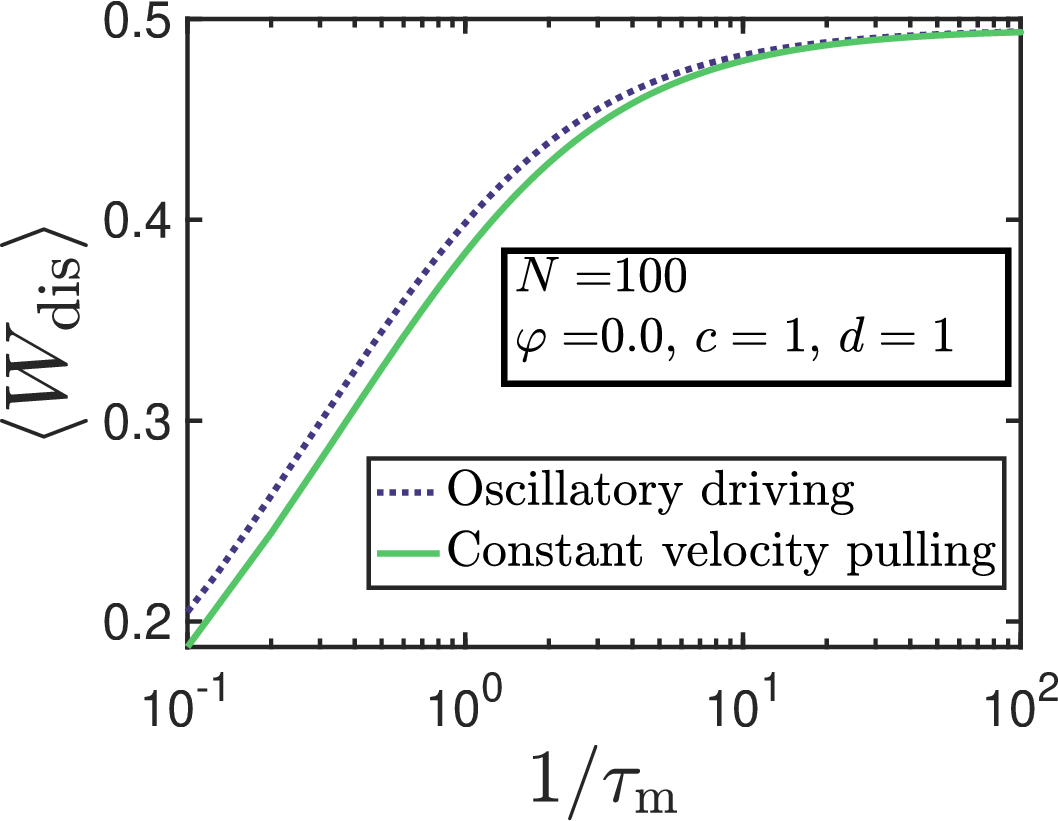}\\
(a)\\
\includegraphics[width=3.in,height=!]{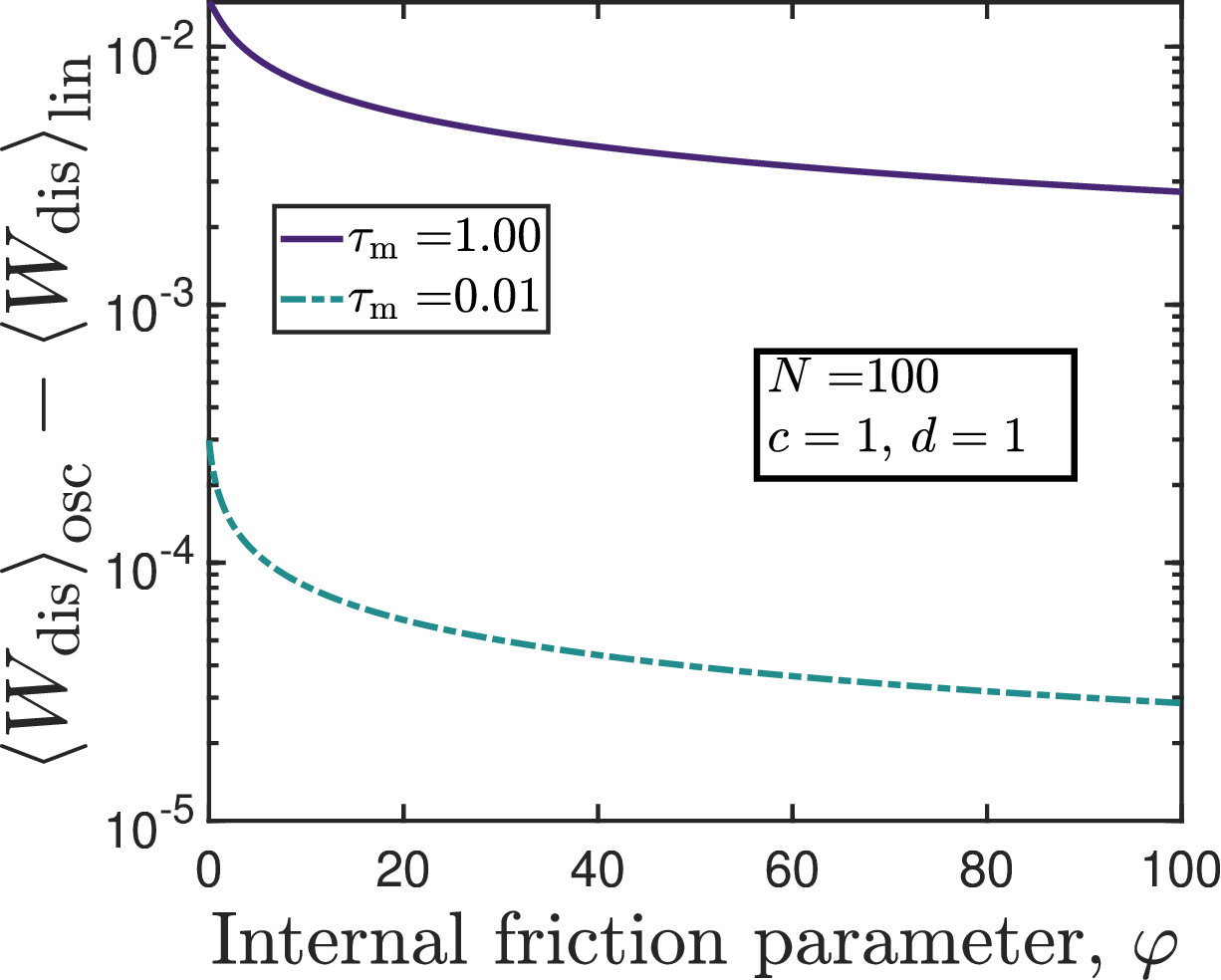}\\
(b)\\
\end{tabular}
\end{center}
\caption{(a) Average dissipated work as a function of protocol duration ($\tau_{\text{m}}$) for a bead-spring-dashpot chain with $N=100$ springs driven under the linear and symmetric protocols (b) Difference between the work dissipated in the two protocols as a function of the internal friction parameter, for the same value of the protocol duration.}
\label{fig:prot_compare}
\end{figure}

\begin{figure}[t]
\begin{center}
\begin{tabular}{c}
\includegraphics[width=3.in,height=!]{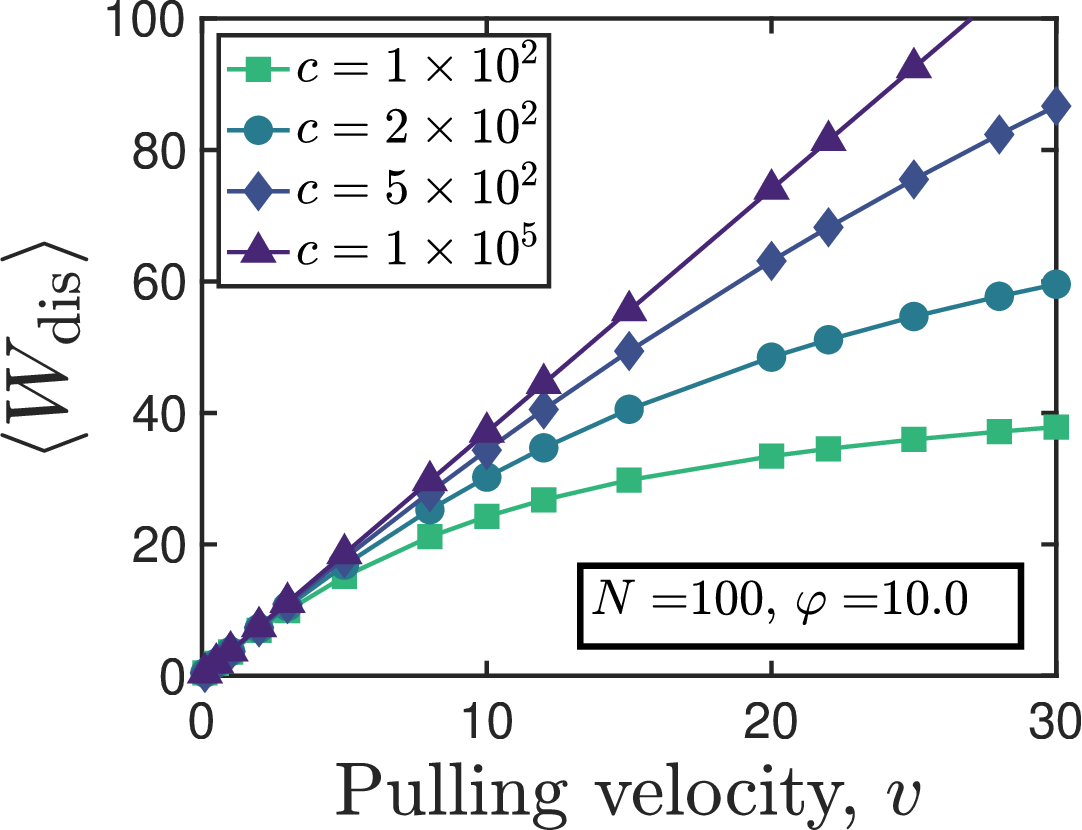}\\
(a)\\
\includegraphics[width=3.in,height=!]{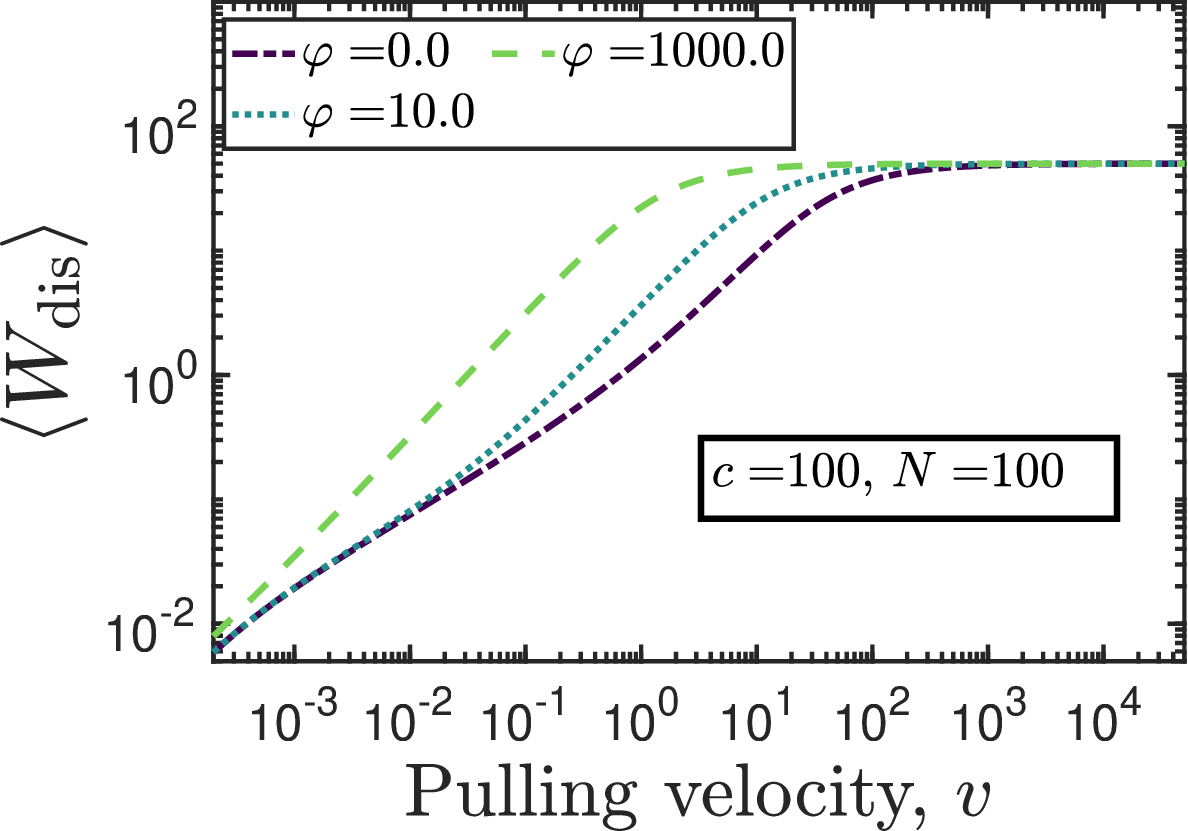}\\
(b)
\end{tabular}
\end{center}
\caption{Average dissipated work as a function of pulling velocity for a chain with internal friction, pulled over a constant dimensionless distance of $d=1$, over (a) a moderate range of pulling velocities, at a fixed internal friction parameter ($\varphi=10$) and various values of the steering spring stiffness, and (b) a larger range of pulling velocities, at a fixed steering spring stiffness ($c=100$) and various values of the internal friction parameter.}
\label{fig:pull_dep_c}
\end{figure}

\begin{figure}[t]
\begin{center}
\begin{tabular}{c}
\includegraphics[width=3.in,height=!]{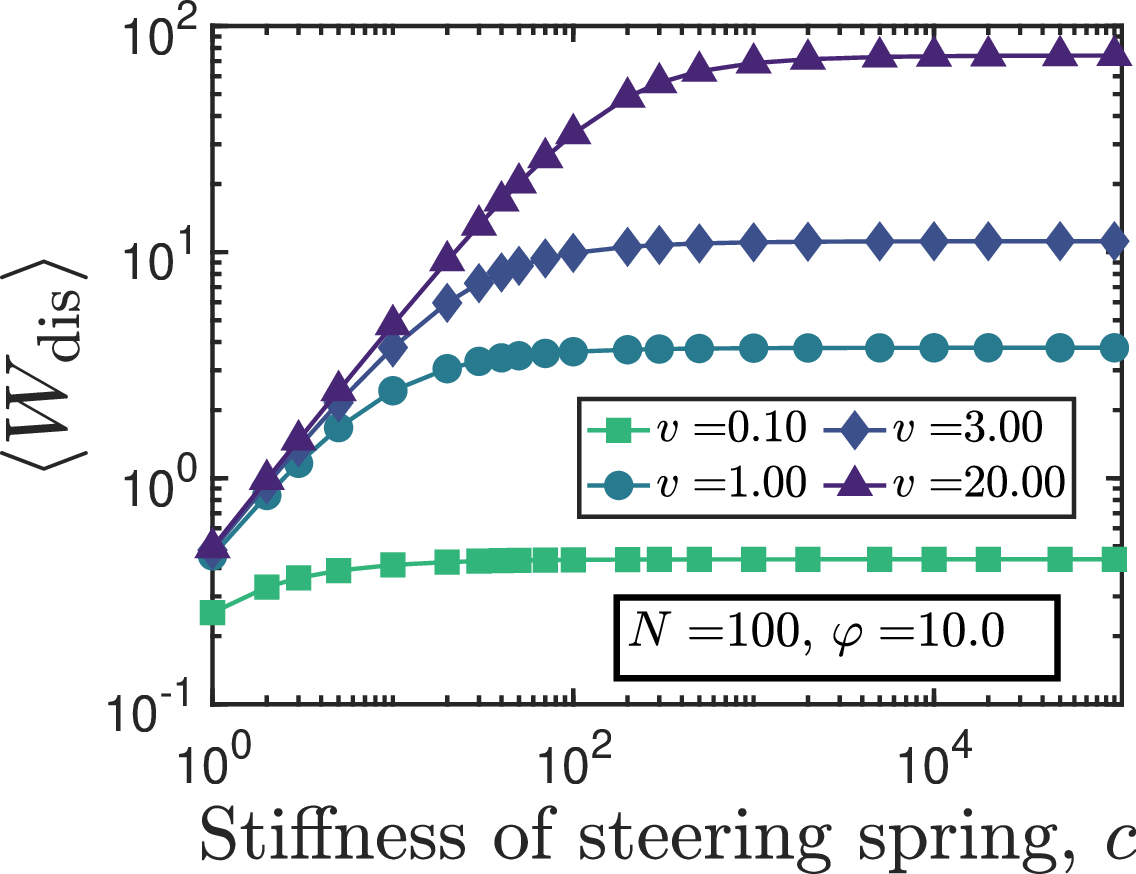}\\
(a)\\
\includegraphics[width=3.in,height=!]{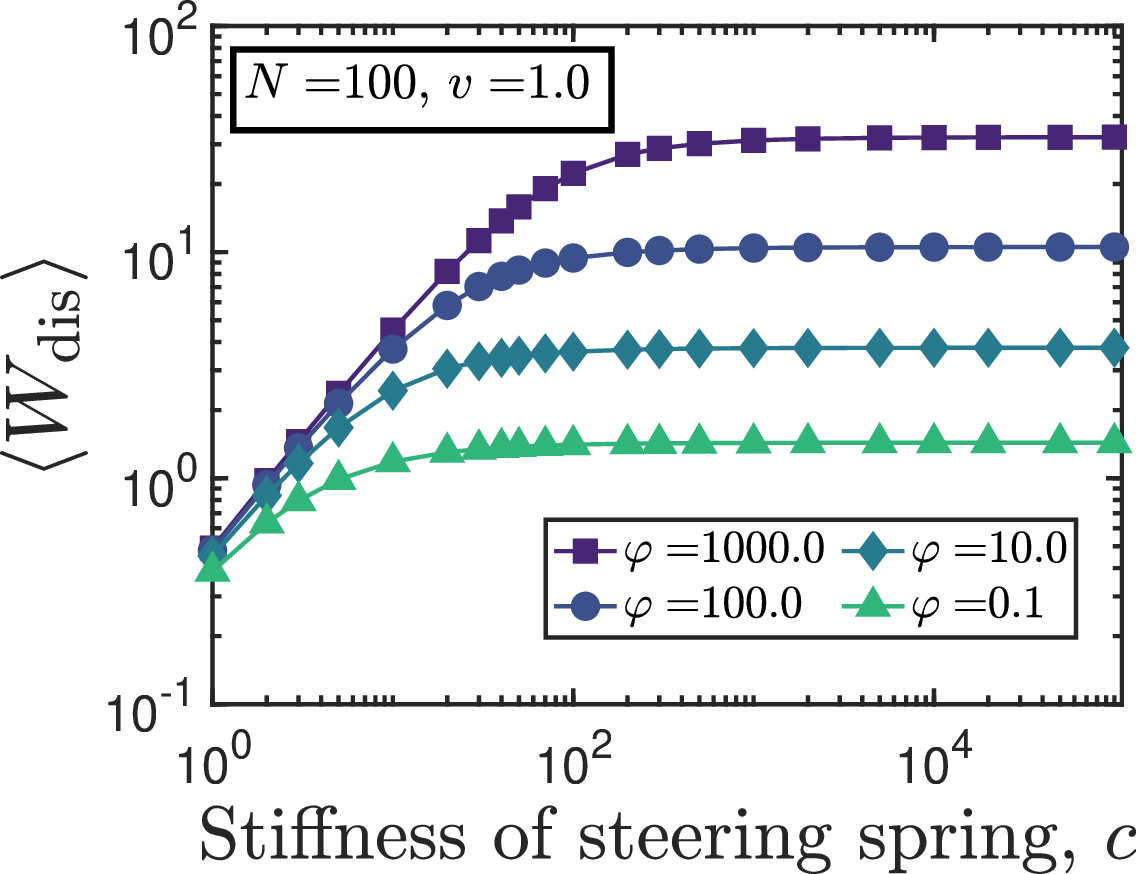}\\
(b)\\
\end{tabular}
\end{center}
\caption{Dissipated work as a function of the steering spring stiffness, for a bead-spring-dashpot chain (a) of fixed internal friction parameter ($\varphi=10$) driven at various pulling velocities and (b) driven at a fixed pulling velocity ($v=1.0$), possessing various values of the internal friction parameter.}
\label{fig:plateau_c}
\end{figure}

 \begin{figure}[t]
\centering
\includegraphics[width=80mm]{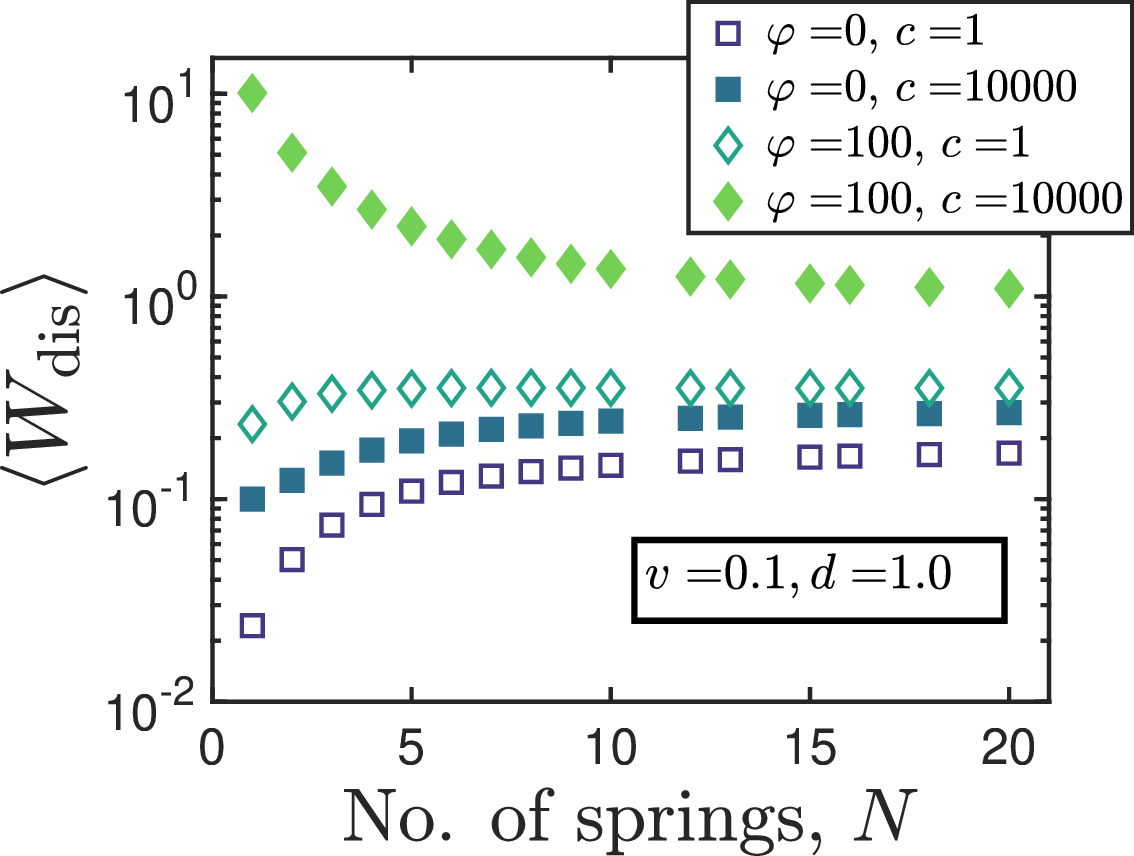}  
\caption{Dissipated work as a function of chain length ($N$) at two values of the trap stiffness, measured for bead-spring-dashpot chains with and without internal friction, pulled at a constant velocity of $v=0.1$ over a distance $d=1.0$.}
\label{fig:c_v_vphi}
\end{figure}

\begin{figure}[t]
\centering
\includegraphics[width=80mm]{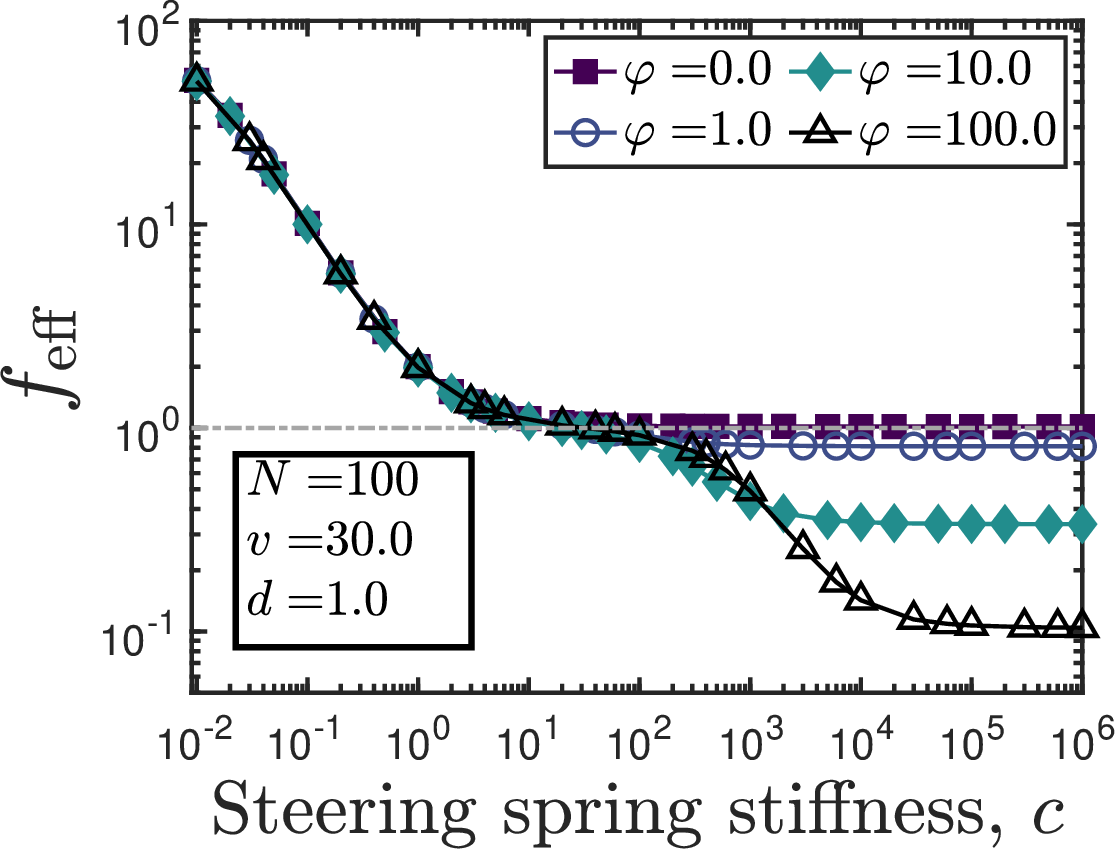}  
\caption{Plot of $f_{\text{eff}}$ [defined in eq.~(\ref{eq:f_eff_def})] as a function of the steering spring stiffness, for a bead-spring-dashpot chain with $N=100$, pulled at a velocity of $v=30$ over a distance $d=1$.}
\label{fig:f_plot}
\end{figure}

 The probability distribution of work done, $P(W)$, is a Gaussian whose expression is given by
 \begin{align}\label{eq:prob_work}
 P(W)=\dfrac{1}{\sqrt{2\pi\sigma^2}}e^{-\left(W-\left<W\right>\right)^2/2\sigma^2}
 \end{align}
 which implies that the Jarzynski equality [cf. Eq.~(\ref{eq:jeq_ori})], is trivially satisfied for this system~\cite{Jarzynski1997,Speck2004,Kailasham2020}.
 
The mean of the Gaussian is given by eq.~(\ref{eq:w_components}), where the free energy difference in transitioning between the initial state (at $\tau=0$) and the final state ($\tau=\tau_{\text{m}}$) is the same for both the protocols considered in this manuscript, i.e., 
\begin{align}
\Delta{F}&={F}\left[\alpha(\tau_{\text{m}})\right]-{F}\left[\alpha(0)\right]\nonumber\\[5pt]
&=\dfrac{cd}{2}\left[\alpha(\tau_{\text{m}})+\alpha(0)\right]\left\{1-c\left[\bm{A}^{-1}\right]_{NN}\right\}
\end{align}
The dissipation for the two protocols, however, are different as evident from eq.~(\ref{eq:wdis_const_vel}) and eq.~(\ref{eq:wdis_o_num}). 

Fig.~\ref{fig:prob_w} illustrates the probability distribution given by eq.~(\ref{eq:prob_work}) for a variety of parameters. For a quasistatic process, the average work done in the pulling process is equal to the free energy change (by definition~\cite{Callen1985}), and there is no associated dissipation. At finite rates of driving however, there is dissipation incurred even in the absence of internal friction, because the beads have to be pulled against the frictional resistance offered by the solvent. With the inclusion of internal friction, the dissipation (width of the distribution) increases for the same value of the pulling velocity. Higher the pulling velocity (or frequency), larger is the dissipation. A comparison between the dissipation incurred in the two protocols is provided by Fig.~\ref{fig:prot_compare}.

Fig.~\ref{fig:prot_compare}~(a) illustrates that for the same duration of the pulling protocol, $\tau_{\text{m}}$, the work dissipated in oscillatory driving of a Rouse chain (without internal friction) is greater than that incurred in constant velocity pulling, in agreement with the findings of~\citet{Speck2005}.  The difference between the two protocols vanishes in the limit of slow pulling ($\tau_{\text{m}}\to0$). Fig~\ref{fig:prot_compare}~(b) shows that the difference in dissipation due to the two protocols measured for the same protocol duration, decreases as a function of the internal friction parameter. 

 \subsection{\label{sec:a_res_disc} Constant velocity pulling (linear protocol)}
 
The formal expression given by eq.~\ref{eq:wdis_const_vel}, written entirely in terms of the matrix elements, does not provide an insight into the dependence of the dissipated work on factors such as the stiffness of the steering spring, the size of the polymer chain ($N$), or how the total dissipation relates to the damping coefficient of a single dashpot. We attempt to address these questions through numerical calculations.

Firstly, we note from Fig.~\ref{fig:pull_dep_c}~(a) that for a fixed chain size, pulling distance and dashpot damping coefficient, the dependence of dissipated work on the pulling velocity changes from being non-linear to linear with an increase in the stiffness of the steering spring. Additionally, Fig.~\ref{fig:pull_dep_c}~(b) illustrates that for a given value of the steering spring stiffness, the dissipated work attains a constant value in the limit of high pulling velocities. Increasing the internal friction parameter results in the attainment of the asymptotic dissipation value at smaller values of the pulling velocity.

The variation of dissipation as a function of the steering trap stiffness is shown in Fig.~\ref{fig:plateau_c}. For a fixed value of the chain length and the pulling distance, the dependence of dissipation on the steering spring stiffness vanishes at large enough values of the latter, plateauing to a constant value. The spring stiffness at which the dissipation attains a constant value appears to be a function of both the pulling velocity and the internal friction coefficient. With an increase of either the internal friction coefficient or the pulling velocity, the plateau value of dissipation is reached at larger values of the steering spring stiffness.

\begin{figure}[t]
\begin{center}
\begin{tabular}{c}
\includegraphics[width=3.3in,height=!]{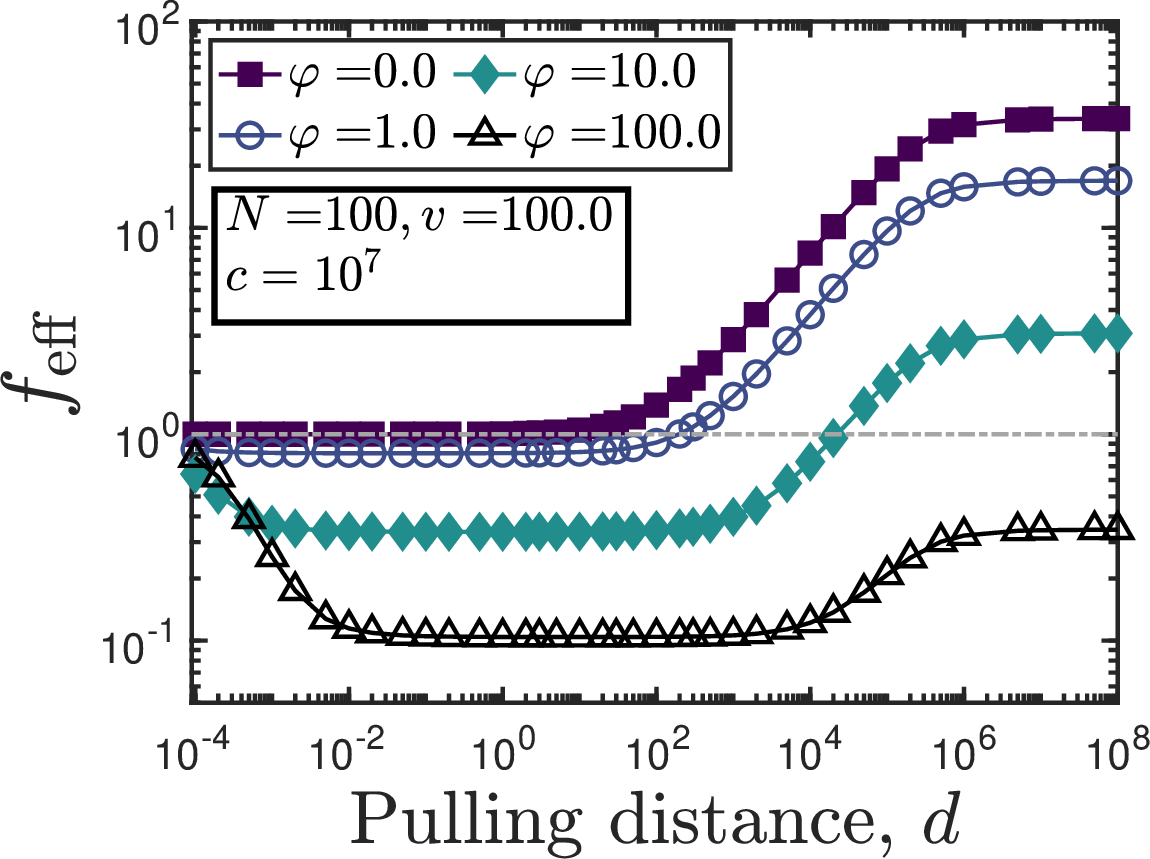}\\
(a)\\
\includegraphics[width=3.3in,height=!]{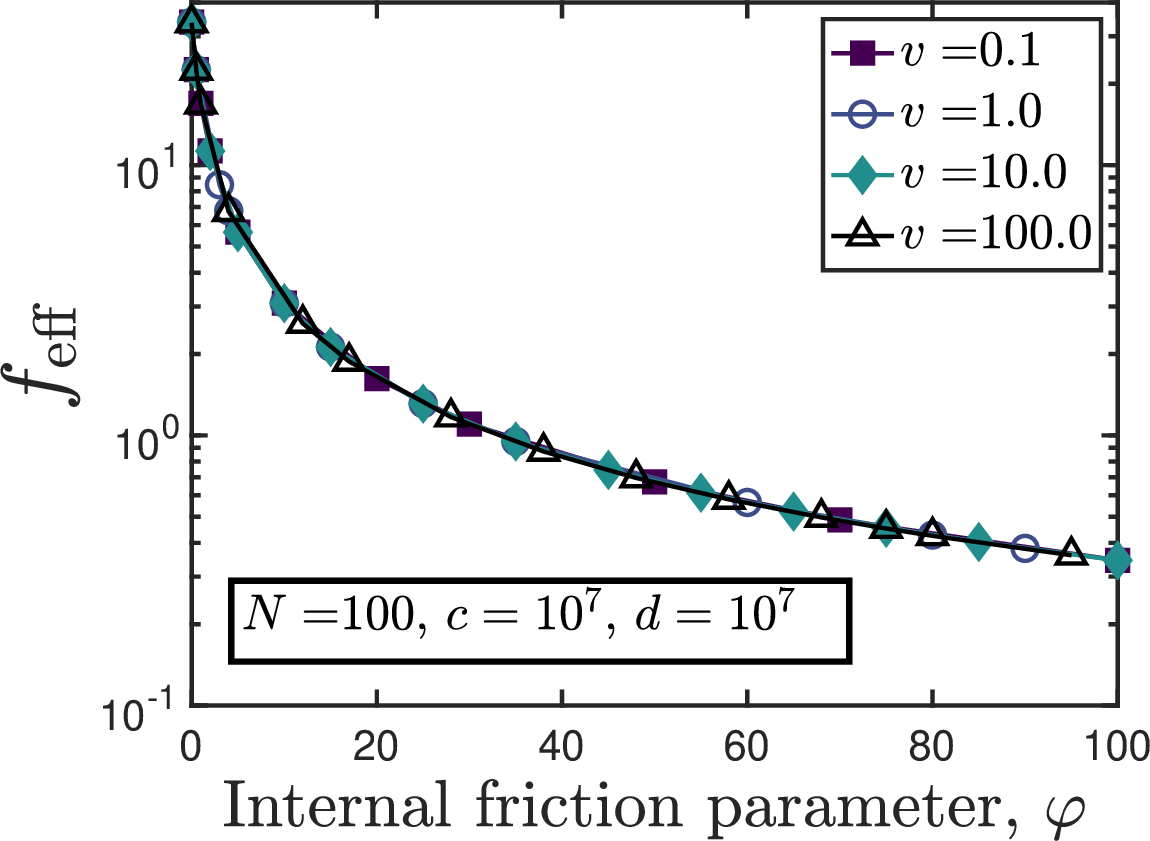}\\
(b)\\
\end{tabular}
\end{center}
\caption{Variation of $f_{\text{eff}}$ [defined in eq.~(\ref{eq:f_eff_def})] as a function of (a) pulling distance in the limit of high trap stiffness and (b) internal friction parameter in the limit of large pulling distance and high trap stiffness, for a bead-spring-dashpot chain with $N=100$ springs.}
\label{fig:f_eff_dep}
\end{figure}

\begin{figure}[t]
\begin{center}
\begin{tabular}{c}
\includegraphics[width=3.in,height=!]{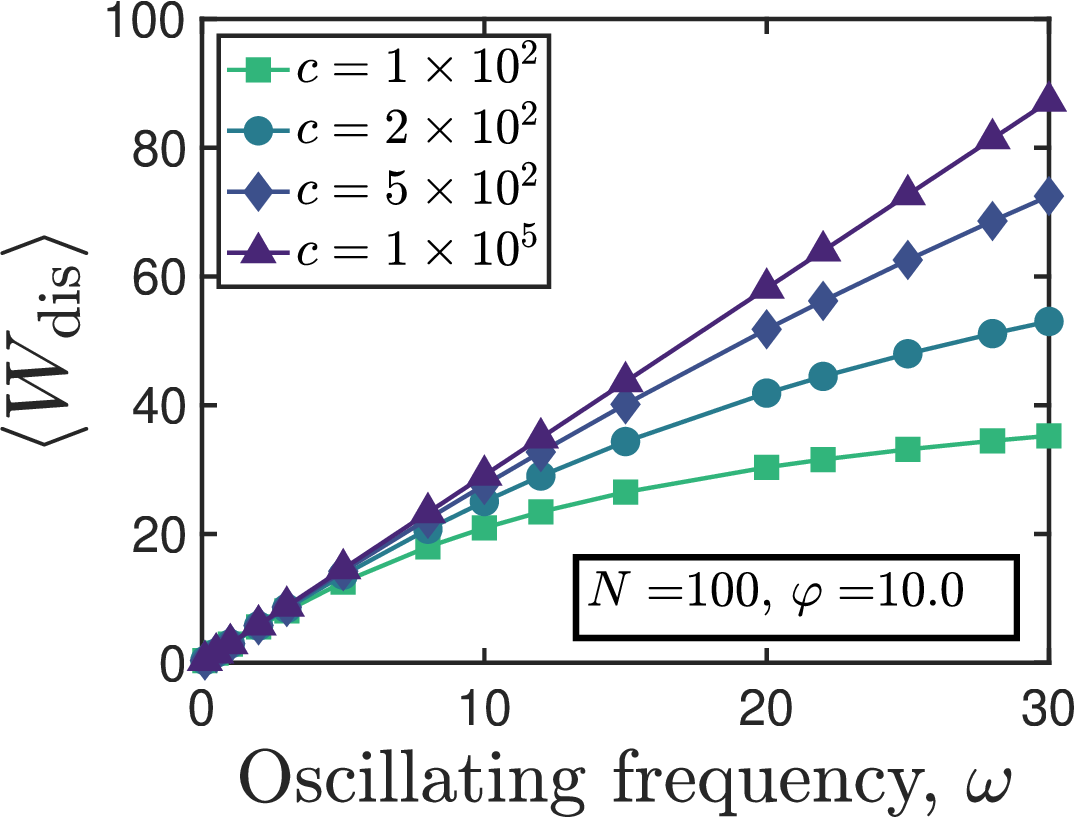}\\
(a)\\
\includegraphics[width=3.in,height=!]{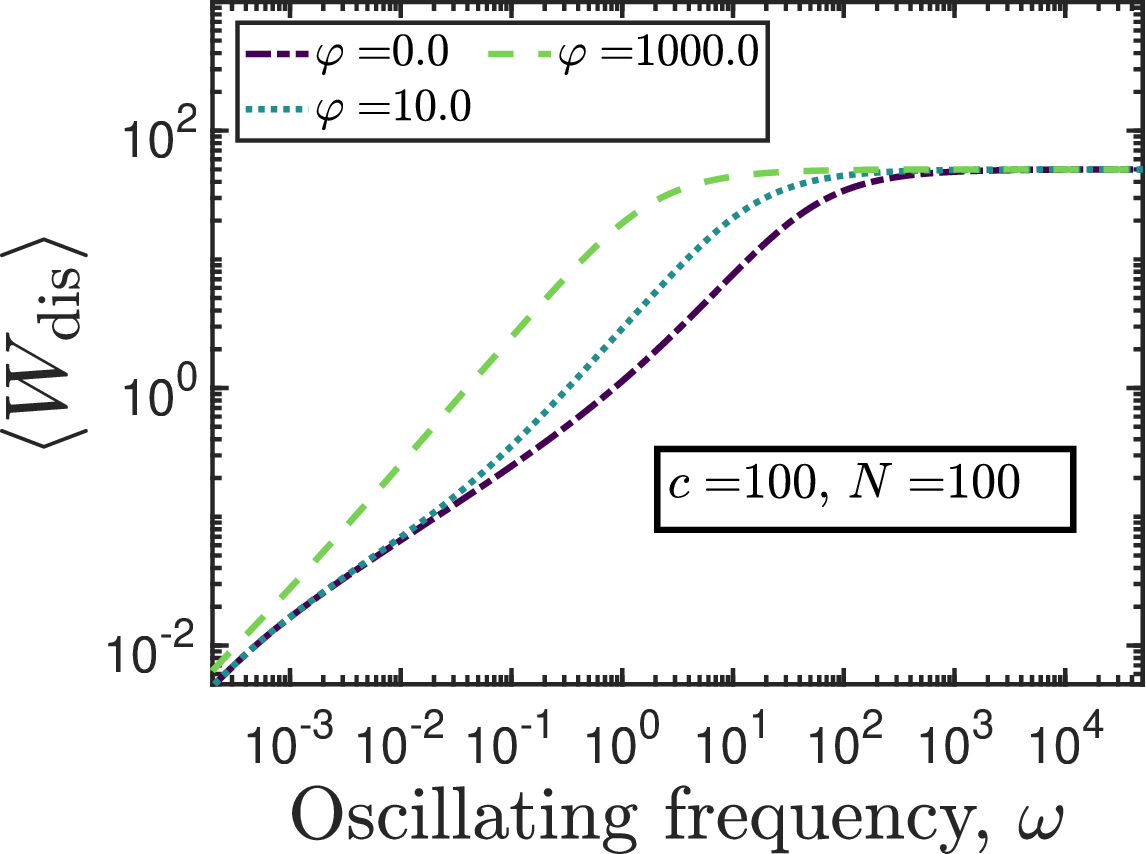}\\
(b)
\end{tabular}
\end{center}
\caption{Average dissipated work as a function of oscillating frequency for a chain with internal friction, pulled over a constant dimensionless distance of $d=1$, over (a) a moderate range of driving frequencies, at a fixed internal friction parameter ($\varphi=10$) and various values of the steering spring stiffness, and (b) a larger range of driving frequencies, at a fixed steering spring stiffness ($c=100$) and various values of the internal friction parameter.}
\label{fig:omega_dep_c}
\end{figure}

\begin{figure}[t]
\begin{center}
\begin{tabular}{c}
\includegraphics[width=3.in,height=!]{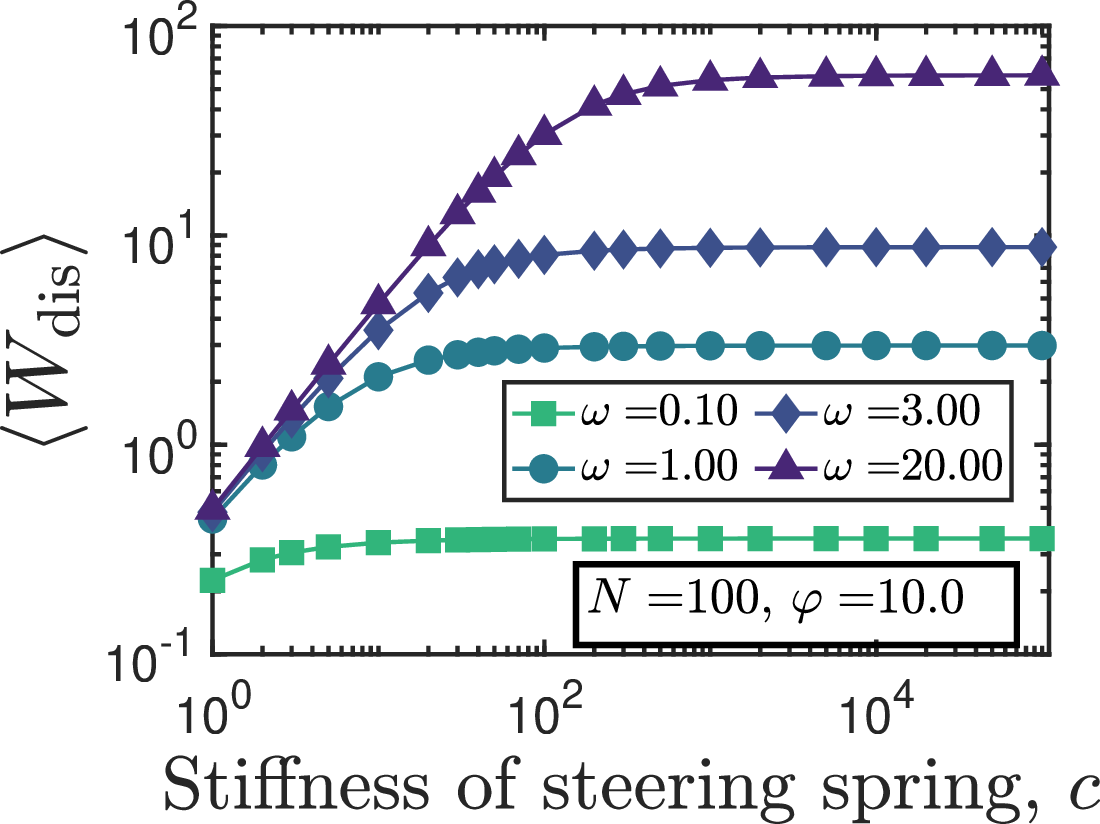}\\
(a)\\
\includegraphics[width=3.in,height=!]{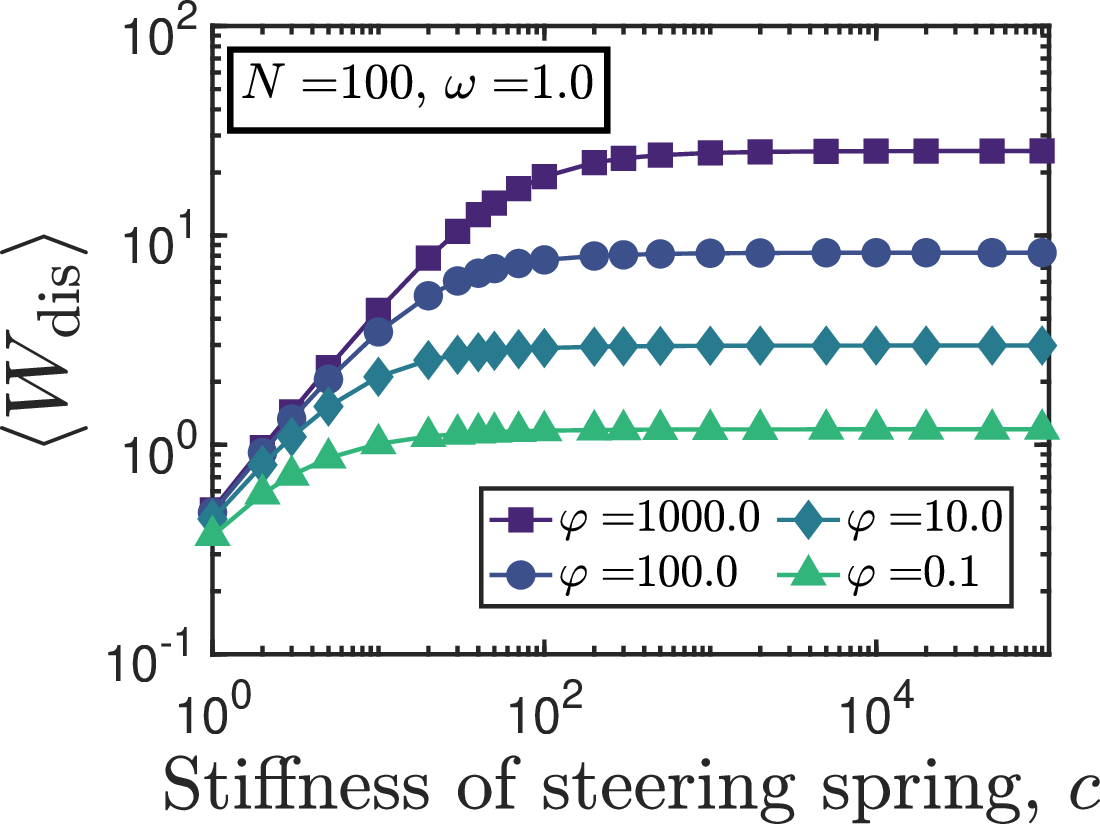}\\
(b)\\
\end{tabular}
\end{center}
\caption{Dissipated work as a function of the steering spring stiffness, for a bead-spring-dashpot chain (a) of fixed internal friction parameter ($\varphi=10$) driven at various frequencies and (b) driven at a fixed frequency ($\omega=1.0$), possessing various values of the internal friction parameter.}
\label{fig:plateau_c_oscill}
\end{figure}

 \begin{figure}[t]
\centering
\includegraphics[width=80mm]{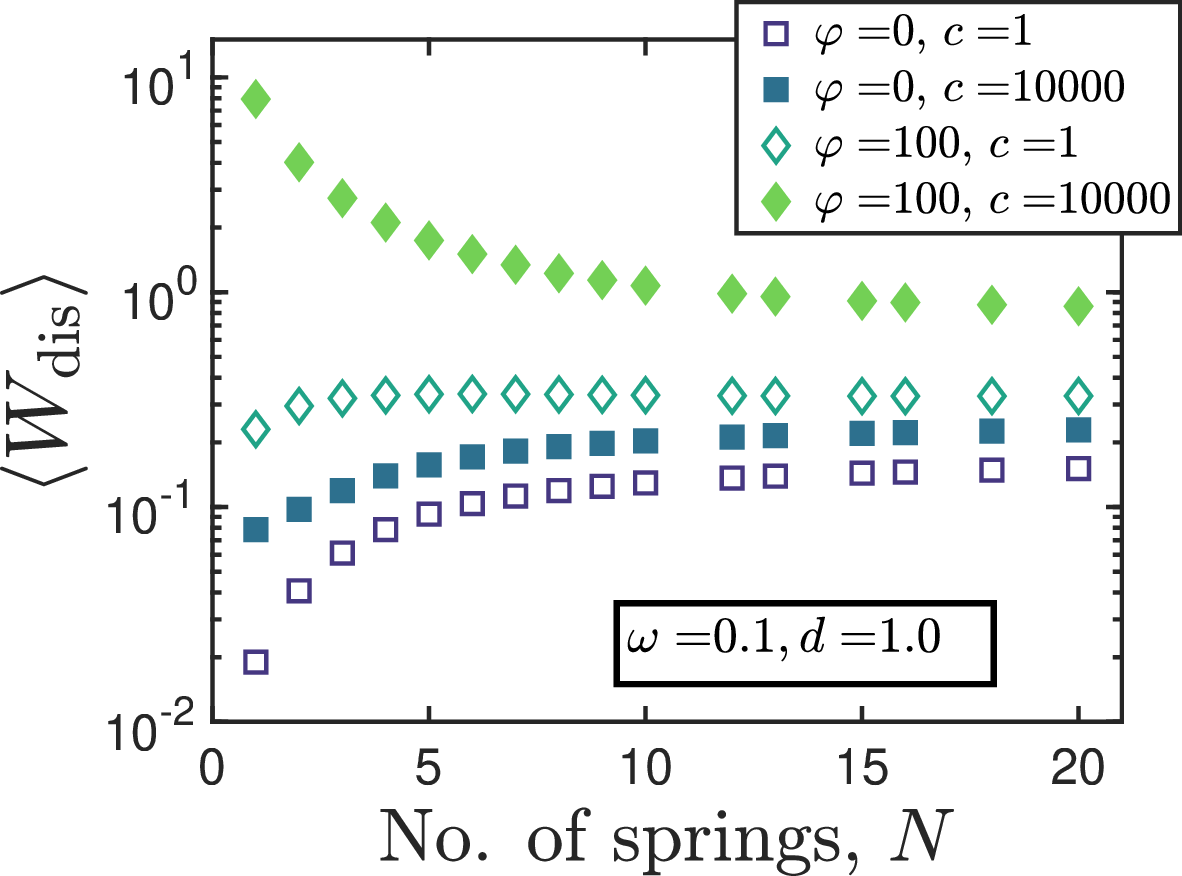}  
\caption{Dissipated work as a function of chain length ($N$) at two values of the trap stiffness, measured for bead-spring-dashpot chains with and without internal friction, driven at a constant frequency of $\omega=0.1$ over a distance $d=1.0$.}
\label{fig:c_omg_vphi}
\end{figure}

A natural question to consider is how the dissipation varies as a function of the chain length? i.e., given a bead-spring-dashpot unit with a particular value of the spring stiffness and dashpot damping coefficient, how will the dissipation incurred in pulling such a setup change as more bead-spring-dashpots are added in series. The answer appears to be highly nontrivial, and dependent on the stiffness of the steering spring, as discussed below.

Fig.~\ref{fig:c_v_vphi} illustrates the dissipated work as a function of the pulling velocity for bead-spring-dashpot chains of various lengths, for the case of a soft and stiff steering spring. In the absence of internal friction ($\varphi=0$), the dissipation at each pulling velocity increases with chain length before ceasing to increase further, for both soft and stiff steering springs. The same trend is observed in the presence of internal friction, for bead-spring-dashpots pulled using a soft spring. This is straight forward, in that more work is dissipated in pulling a chain with two spring-dashpots, say, as compared to a chain with ten spring-dashpots. We term this situation as being \textit{cooperative}, as the spring-dashpots contribute additively towards the total dissipation. Note that the term additive is used in a loose sense because it is not known precisely if the contribution to the total dissipation from each spring-dashpot is exactly additive or some other function.

For chains with internal friction being pulled using a stiff steering spring, however, the dissipation incurred at each pulling velocity \textit{decreases} with an increase in chain length. This appears surprising, because, it implies that pulling a chain with two spring-dashpots dissipates more work than one with ten spring-dashpots. We term this situation as being \textit{anti-cooperative}, since the dissipation incurred by several dashpots is lower than a single spring-dashpot being pulled. This trend remains conserved even when the $total$ work (not shown) and not just the dissipated work is considered. A transition from cooperative to anti-cooperative response is not observed in the absence of internal friction, however.

The usage of a harmonic trap with large stiffness to minimize the dissipation incurred in pulling a colloidal particle over an undulating energy landscape has been reported by Sivak and coworkers~\cite{Blaber2022,Blaber2022a}. Specifically, they aim to develop a two-dimensional protocol for minimizing the dissipation associated with the above process, by tuning the position of the trap centre and the stiffness of harmonic confinement with time. They find that the protocol resulting minimal dissipation requires that the trap stiffness increase as the colloid crosses the barriers. Additionally, it is found that higher values of the initial trap stiffness result in lower values of the dissipation and hence higher efficiency.

The dissipative response of a series of spring-dashpots being pulled at a constant velocity is clearly dictated by the stiffness of the steering spring. We next examine the value of the spring-stiffness at which the dissipative response transitions from being cooperative to anti-cooperative, and its dependence on the internal friction coefficient. We introduce the quantity $f_{\text{eff}}(N;\varphi,v,c,d)$ which is a ratio of the work dissipated by a chain of $N$ bead-spring-dashpots to a single bead-spring-dashpot, at fixed values of the pulling velocity, pulling distance, the dashpot damping coefficient and the steering spring stiffness.
\begin{align}\label{eq:f_eff_def}
f_{\text{eff}}(N;\varphi,v,c,d)=\dfrac{\left<W_{\text{dis}}\right>_{N}}{\left<W_{\text{dis}}\right>_{N=1}}\Bigg\vert_{\varphi,v,c,d}
\end{align}
A value of $f_{\text{eff}}>1$ suggests cooperative behavior, while $f_{\text{eff}}<1$ is indicative of an anticooperative response. A single spring-dashpot ($N=1$) subjected to pulling will therefore have a value of $f_{\text{eff}}=1$ by definition, at all values of the steering spring stiffness. In the limit of large pulling spring stiffness, the following relationship may be written between the work dissipated in a bead-spring-dashpot chain with $N$ spring-dashpots to that dissipated by a single spring-dashpot (cf. eq.~(\ref{eq:vel_N1})): 
\begin{align}
\lim_{c\to\infty}\left<W_{\text{dis}}\right>_{N}=f_{\text{eff}}(1+\varphi)vd
\end{align}

Fig.~\ref{fig:f_plot} illustrates the variation of $f_{\text{eff}}$ as a function of the steering spring stiffness for various values of the internal friction coefficient. We observe that the steering spring stiffness at which the transition from cooperative to anti-cooperative response occurs is a function of the pulling velocity and distance. In the limit of large trap stiffness, $f_{\text{eff}}$ attains a plateau that is dependent on the internal friction parameter, at constant values of the pulling velocity and distance.

Fig.~\ref{fig:f_eff_dep}~(a) illustrates the variation of $f_{\text{eff}}$ as a function of the pulling distance, in the limit of large steering spring stiffness, for chains with various values of the internal friction parameter. At infinitesimally small values of the pulling distance, the value of $f_{\text{eff}}$ is unity for all various of the internal friction parameter. In the limit of large pulling distance, $f_{\text{eff}}$ attains a plateau value that is dependent on the internal friction parameter. For large values of the pulling distance and trap stiffness, Fig.~\ref{fig:f_eff_dep}~(b) illustrates that $f_{\text{eff}}$ is independent of the pulling velocity, and is only a function of the internal friction parameter.

Owing to this dependence of $f_{\text{eff}}$ on the internal friction coefficient, it cannot be estimated $a\,priori$. This is in direct contrast with the single mode spring-dashpot, for which $f_{\text{eff}}=1$ irrespective of the internal friction coefficient. Consequently, it is not possible in general to establish a relationship between the total dissipation in pulling a bead-spring-dashpot chain at constant velocity and the damping coefficient of a single dashpot. The specific case of a single-mode spring dashpot is an exception, and is relatively easier to analyze, as discussed in ref.~\citenum{Kailasham2020}. 

 \subsection{\label{sec:b_res_disc} Oscillatory driving (symmetric protocol)}

\begin{figure}[t]
\centering
\includegraphics[width=80mm]{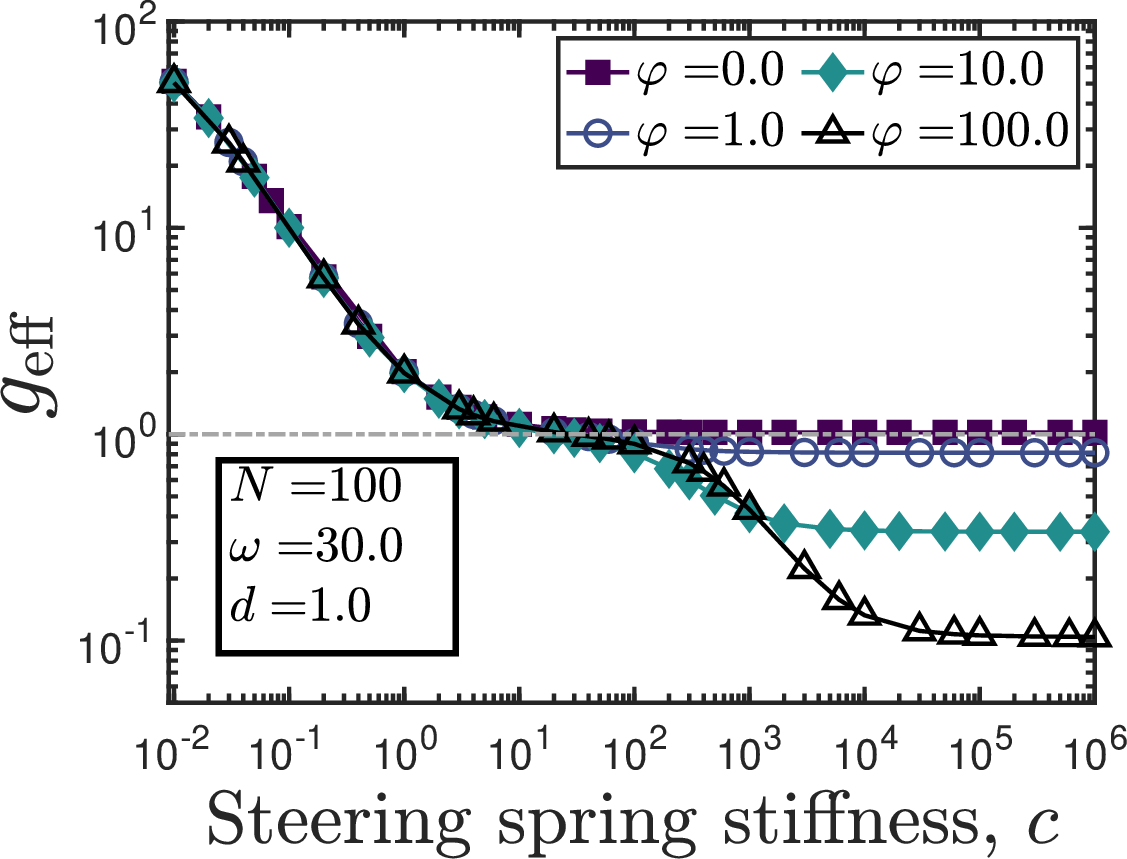}  
\caption{Plot of $g_{\text{eff}}$ [defined in eq.~(\ref{eq:g_eff_def})] as a function of the steering spring stiffness, for a bead-spring-dashpot chain with $N=100$, driven at a frequency of $\omega=30$ over a distance $d=1$.}
\label{fig:g_plot}
\end{figure}

\begin{figure}[t]
\begin{center}
\begin{tabular}{c}
\includegraphics[width=3.3in,height=!]{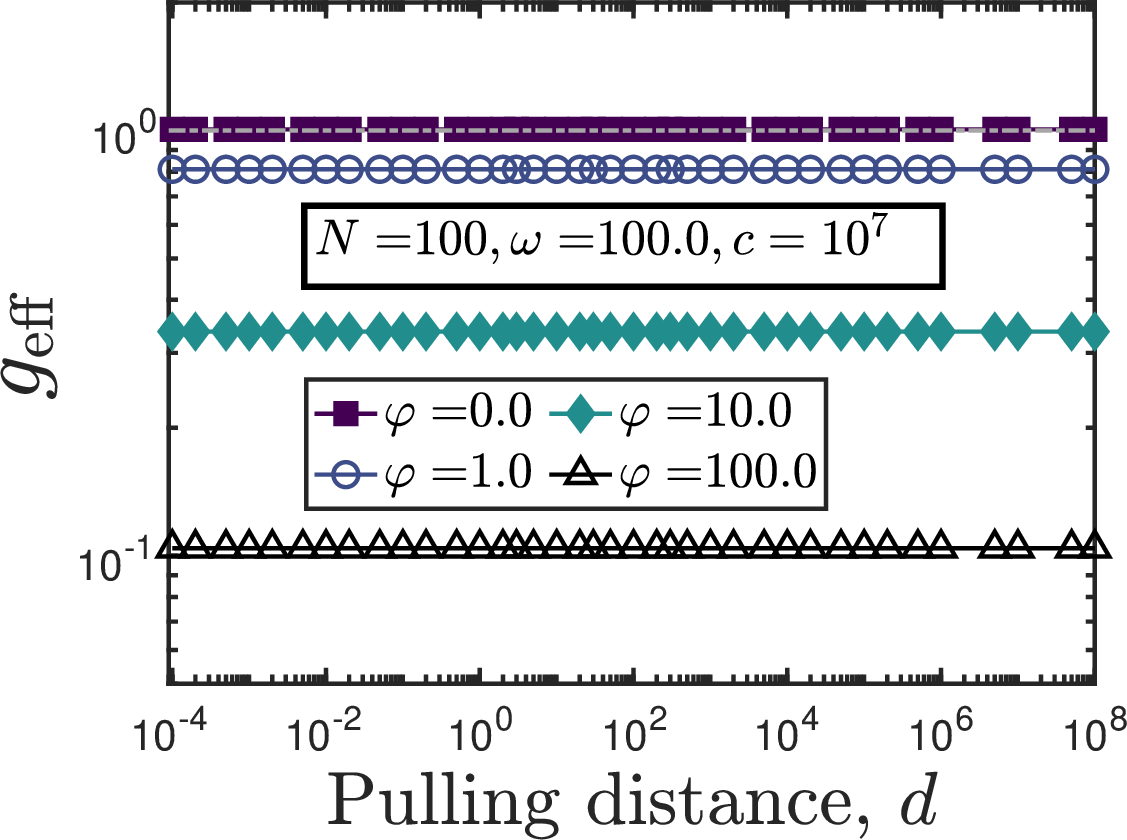}\\
(a)\\
\includegraphics[width=3.3in,height=!]{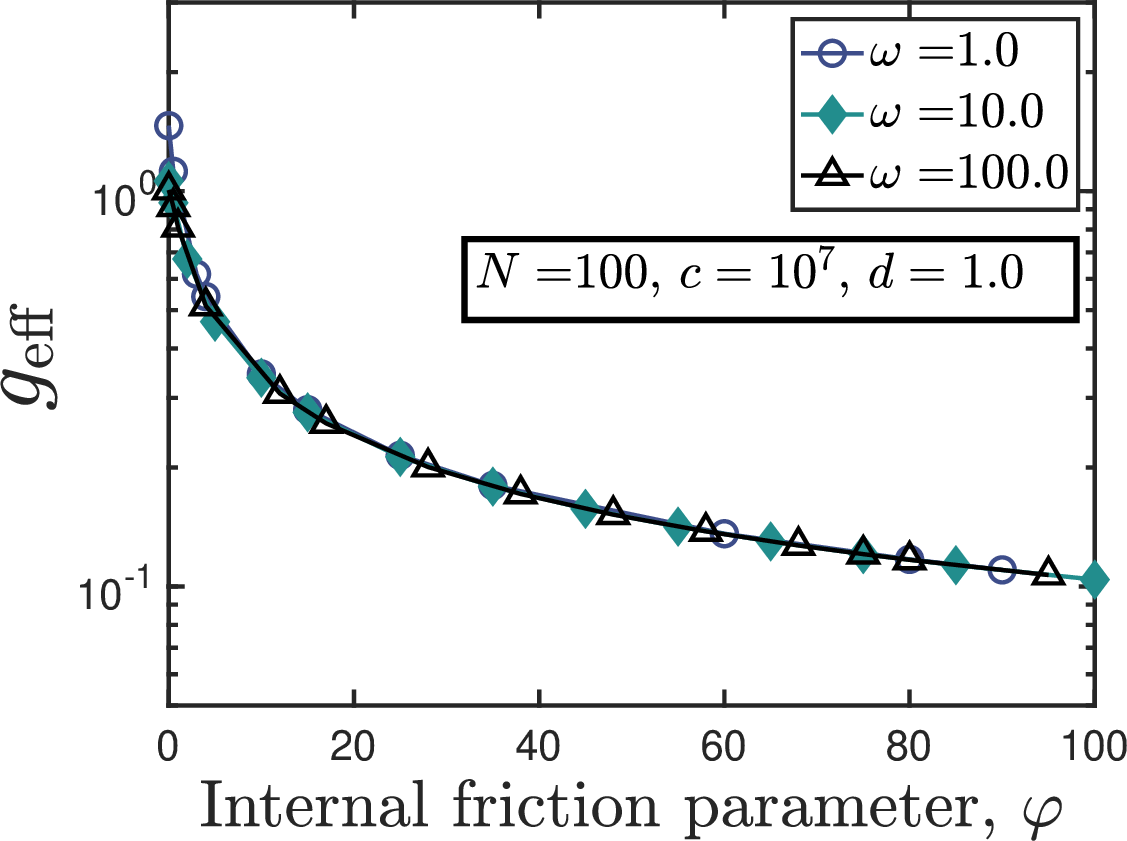}\\
(b)\\
\end{tabular}
\end{center}
\caption{Variation of $g_{\text{eff}}$ [defined in eq.~(\ref{eq:g_eff_def})] as a function of (a) pulling distance and (b) internal friction parameter in the limit of high trap stiffness for a bead-spring-dashpot chain with $N=100$ springs.}
\label{fig:g_eff_dep}
\end{figure}

 \begin{figure}[t]
\begin{center}
\begin{tabular}{c}
\includegraphics[width=3.3in,height=!]{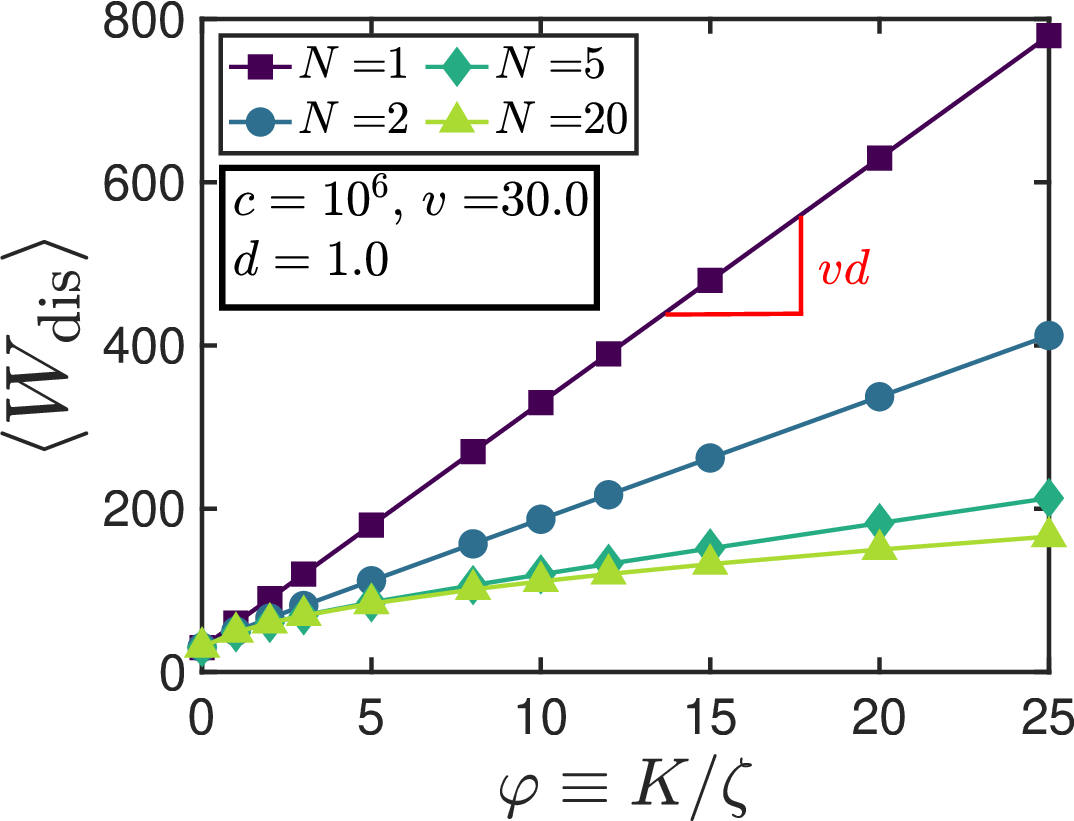}\\
(a)\\
\includegraphics[width=3.3in,height=!]{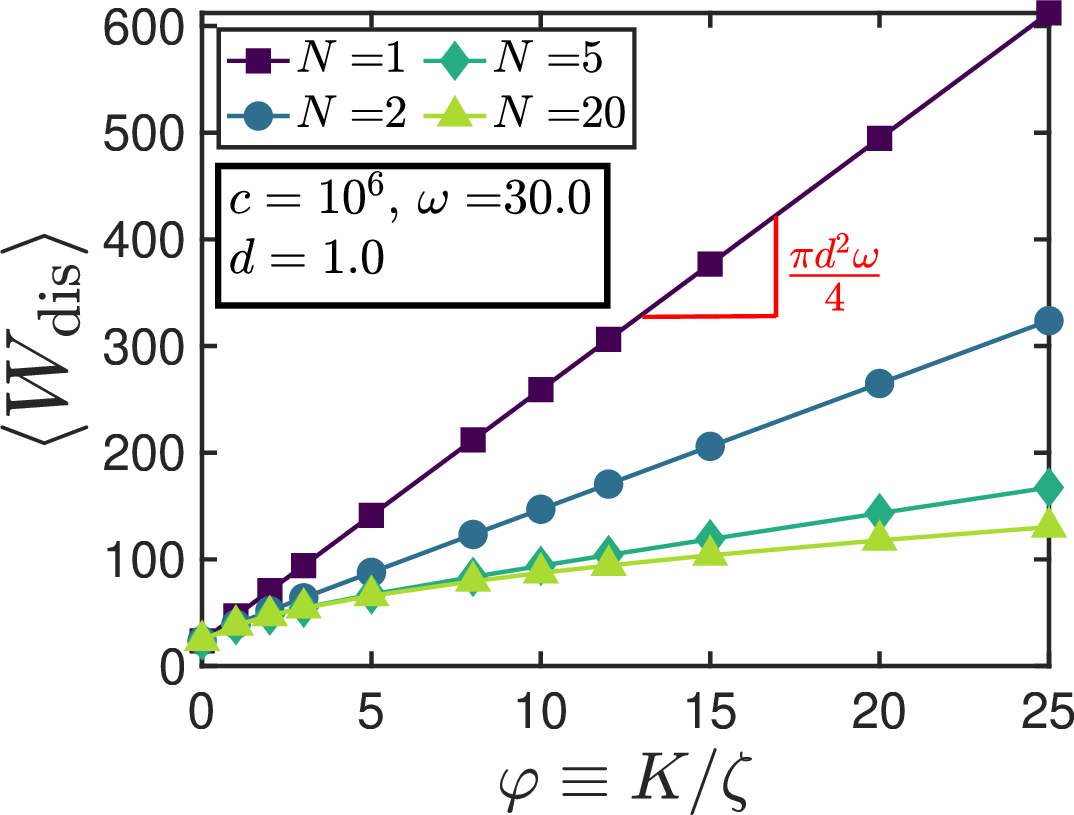}\\
(b)\\
\end{tabular}
\end{center}
\caption{Variation of dissipated work as a function of the internal friction parameter for (a) constant velocity pulling (linear protocol) and (b) oscillatory driving (symmetric) protocol, for chains of various lengths. The pulling trap stiffness and pulling distance are fixed at $c=10^6$ and $d=1$ for both the cases.}
\label{fig:nlin_vphi}
\end{figure}

The variation of the dissipation incurred in oscillatory driving as a function of frequency shows a near-identical trend to that observed in constant velocity pulling. As illustrated in Fig.~\ref{fig:omega_dep_c}~(a), over a moderate range of driving frequencies, the scaling of dissipation with respect to the frequency goes from being non-linear to linear as the trap stiffness is increased, keeping all other parameters fixed. Furthermore, Fig.~\ref{fig:omega_dep_c}~(b) shows that the dissipation incurred in the large frequency limit is a function of the trap stiffness and is independent of the internal friction parameter. The inclusion of internal friction results in the attainment of the asymptotic dissipation value at lower frequencies.

Fig.~\ref{fig:plateau_c_oscill} illustrates that the dissipated work increases with the steering trap stiffness, before attaining a constant value that depends on the frequency of the driving and the internal friction parameter. This trend is identical to that observed for the case of constant velocity pulling.

In addressing the question of how the dissipation in oscillatory driving scales with the chain length, we encounter a similar trend as observed for the case of constant velocity pulling. As seen from Fig.~\ref{fig:c_omg_vphi}, for chains without internal friction, the dissipation increases with the number of springs in the chain, irrespective of the stiffness of the steering spring, and saturates to a constant value. With the inclusion of internal friction, however, the steering spring stiffness determines if the dissipation increases or decreases with the number of spring-dashpots in the chain.

The transition from cooperative to anti-cooperative response is observed in the case of oscillatory driving as well, and we introduce the metric $g_{\text{eff}}(N;\varphi,\omega,c,d)$ to quantify this phenomenon, which is defined as the ratio of work done dissipated in driving a $N$-spring-dashpot chain with internal friction parameter $\varphi$ over a distance $d$ at a frequency $\omega$, to that dissipated in driving a single spring-dashpot ($N=1$) at the same  set of parameter values. Mathematically,  
\begin{align}\label{eq:g_eff_def}
g_{\text{eff}}(N;\varphi,\omega,c,d)=\dfrac{\left<W_{\text{dis}}\right>_{N}}{\left<W_{\text{dis}}\right>_{N=1}}\Bigg\vert_{\varphi,\omega,c,d}
\end{align}
A value of $g_{\text{eff}}>1$ suggests cooperative behavior, while $g_{\text{eff}}<1$ is indicative of an anticooperative response. A single spring-dashpot ($N=1$) to oscillatory driving will therefore have a value of $g_{\text{eff}}=1$ by definition, at all values of the steering spring stiffness. In the limit of large pulling spring stiffness, the following relationship may be written between the work dissipated in a bead-spring-dashpot chain with $N$ spring-dashpots to that dissipated by a single spring-dashpot (cf. eq.~(\ref{eq:oscil_large_c})): 
\begin{align}
\lim_{c\to\infty}\left<W_{\text{dis}}\right>_{N}=g_{\text{eff}}\left[\dfrac{\pi d^2\omega}{4}(1+\varphi)\right]
\end{align}

Fig.~\ref{fig:g_plot} shows the variation of $g_{\text{eff}}$ as a function of the steering spring stiffness for bead-spring-dashpot chains with various internal friction parameter values, for fixed values of chain length, pulling distance and driving frequency. The crossover from cooperative ($g_{\text{eff}}>1$) to anti-cooperative response ($g_{\text{eff}}<1$) is observed only in the presence of internal friction, i.e., $\varphi>0$, similar to that observed in constant velocity pulling.

In a major departure from the constant velocity pulling case, however, $g_{\text{eff}}$ is not a function of the pulling distance, as evinced in Fig.~\ref{fig:g_eff_dep}~(a). The limiting value of $g_{\text{eff}}$ at high spring stiffness is then observed to be a function of the internal friction parameter alone, and is independent of the driving frequency, as illustrated in Fig.~\ref{fig:g_eff_dep}~(b).

The dependence of the prefactor $g_{\text{eff}}$ ($f_{\text{eff}}$) on the internal friction parameter in the symmetric (linear) protocol therefore implies that the work dissipated in the process does not scale linearly with the internal friction parameter in the limit of high trap stiffness. 

Fig.~\ref{fig:nlin_vphi} illustrates the scaling of the dissipated work as a function of the internal friction parameter for various values of the chain length. In the limit of high trap stiffness, for both the driving protocols considered in this manuscript, the dissipation goes from being a linear function of $\varphi$ for the $N=1$ case, to a sub-linear function of $\varphi$ as the value of $N$ is increased.

 \section{\label{sec:otl_concl} Outlook and conclusions} 

 We have numerically evaluated the work statistics for a coarse-grained polymer model with internal friction subjected to driving protocols commonly encountered in single-molecule force spectroscopy experiments, namely, constant velocity (linear) and oscillatory (symmetric) pulling. The dependence of the dissipated work with the pulling trap stiffness, the internal friction parameter and the length of the chain are assessed for both the protocols. {Standard Rouse results are recovered upon setting the internal friction parameter to zero in the current model.}
 
 {A major point of departure from Rouse chain behavior is the pulling trap-stiffness-dependent variation of the total dissipation with respect to the number of springs in the chain. For a bead-spring (Rouse) chain, without internal friction, the average work dissipated in driving the chain through a linear or symmetric protocol increases with the number of springs in the chain, keeping all the other parameters fixed, before saturating to a finite value that is independent of $N$. This response is unaltered by the stiffness of the trap used for effecting the driving. For bead-spring-dashpot chains, i.e., in the presence of internal friction, the dissipation increases with the number of springs in the chain, for small values of the pulling trap stiffness. At large enough values of the pulling trap stiffness, however, the trend is reversed, and the dissipation decreases with $N$, keeping all the other parameters fixed. This anti-cooperative response appears to be a distinguishing feature of coarse-grained polymer models with internal friction.}
 
 {A consequence of this stiffness-dependent dissipative response is that it is not possible to specify a straightforward relationship between the work dissipated in pulling a bead-spring-dashpot chain (with $N>1$) and the damping coefficient of a single dashpot. This difficulty does not arise for the simplest case of a single-mode spring-dashpot ($N=1$), for which an explicit expression has been derived~\cite{Kailasham2020} connecting the damping coefficient of the dashpot to the work dissipated in stretching it at constant velocity. It therefore seems that the quantification of internal friction through driven transitions between equilibrium steady states is complicated. A potentially interesting future research direction would be the quantification of IV through transitions between non-equilibrium steady states, such as a polymer with internal friction being made to unravel by increasing the elongational flow rate of the fluid it is suspended in.~\cite{Latinwo2013,Latinwo2014,Latinwo2014a,Zhou2022,Ghosal2016,Sharma2011,Ghosal2016a}}

 \begin{acknowledgments}
R. K. thanks the Indian Institute of Technology Indore for supporting this work through the Young Faculty Research Seed Grant Scheme (Project No. IITI/YFRSG/2024-2025/Phase-VII/01).
 \end{acknowledgments}
 
 \section*{Data Availability Statement}
 
 MATLAB codes used for generating the figures in this work are available freely on GitHub~\cite{Kailasham_MATLAB_2025}.

\appendix*
\begin{widetext}
\section{\label{sec:app_a} Simplifications used in the derivation of work expression}
This section contains the detailed steps pertaining to the simplification of equations that appear in the derivation of the work expression.
\subsection{\label{sec:i_derv} Simplifying the second term on the RHS of eq.~(\ref{eq:mean_work_step2})}
Defining 
\begin{equation}\label{eq:def_i}
I=\int_{0}^{\tau_{\text{m}}}d\tau\dot{\bm{h}}(\tau)\cdot\int_{0}^{\tau}d\tau'\bm{G}(\tau-\tau')\cdot\bm{D}\cdot\boldsymbol{h}(\tau')
\end{equation}
we start by simplifying the term within the square braces on the RHS of eq.~(\ref{eq:def_i}). Noting that
\begin{equation}\label{eq:intermed_1}
\dfrac{d}{d\tau'}\left\{\bm{G}(\tau-\tau')\right\}\cdot\bm{A}^{-1}=\bm{G}(\tau-\tau')\cdot\bm{D},
\end{equation}
and integrating the inner integral in eq.~(\ref{eq:def_i}) by parts, we may write
\begin{align}\label{eq:inner_integ}
&\int_{0}^{\tau}d\tau'\bm{G}(\tau-\tau')\cdot\bm{D}\cdot\boldsymbol{h}(\tau')=\bm{A}^{-1}\cdot\bm{h}(\tau)-\bm{G}(\tau)\cdot\bm{A}^{-1}\cdot\bm{h}(0)-\int_{0}^{\tau}d\tau'\bm{G}(\tau-\tau')\cdot\bm{A}^{-1}\cdot\dot{\bm{h}}(\tau')
\end{align}
Substituting eq.~(\ref{eq:inner_integ}) into eq.~(\ref{eq:def_i}), we obtain
\begin{align}\label{eq:i_final}
I&=\int_{0}^{\tau_{\text{m}}}d\tau\dot{\bm{h}}(\tau)\cdot\bm{A}^{-1}\cdot\bm{h}(\tau)-\int_{0}^{\tau_{\text{m}}}d\tau\dot{\bm{h}}(\tau)\cdot\bm{G}(\tau)\cdot\bm{A}^{-1}\cdot\bm{h}(0)-\int_{0}^{\tau_{\text{m}}}d\tau\int_{0}^{\tau}d\tau'\dot{\bm{h}}(\tau)\cdot\bm{G}(\tau-\tau')\cdot\bm{A}^{-1}\cdot\dot{\bm{h}}(\tau')
\end{align}
which, upon substitution into eq.~(\ref{eq:mean_work_step2}), yields eq.~(\ref{eq:av_work_intermed}).

\subsection{\label{sec:j_derv} Simplifying the second term on the RHS of eq.~(\ref{eq:av_work_intermed})}
Defining
\begin{equation}\label{eq:def_j}
J=\int_{0}^{\tau_{\text{m}}}d\tau{\dot{\bm{h}}(\tau)}\cdot\bm{A}^{-1}\cdot\bm{h}(\tau)
\end{equation}
and integrating by parts, we get
\begin{align}\label{eq:j_eval_1}
J&=\bm{h}(\tau)\cdot\bm{A}^{-1}\cdot\bm{h}(\tau)\bigg\rvert_{0}^{\tau_{\text{m}}}-\int_{0}^{\tau_{\text{m}}}d\tau\dot{\bm{h}}(\tau)\cdot\bm{A}^{-1}\cdot\bm{h}(\tau)=\bm{h}(\tau_{\text{m}})\cdot\bm{A}^{-1}\cdot\bm{h}(\tau_{\text{m}})-\bm{h}(0)\cdot\bm{A}^{-1}\cdot\bm{h}(0)-J
\end{align}
and finally, 
\begin{equation}\label{eq:j_ans}
J=\dfrac{1}{2}\left[\bm{h}(\tau_{\text{m}})\cdot\bm{A}^{-1}\cdot\bm{h}(\tau_{\text{m}})-\bm{h}(0)\cdot\bm{A}^{-1}\cdot\bm{h}(0)\right]
\end{equation}
Substituting eq.~(\ref{eq:j_ans}) into eq.~(\ref{eq:av_work_intermed}) and simplifying results in eq.~(\ref{eq:w_av_simp2}).

\subsection{\label{sec:id_derv}Simplifying the second term on the RHS of eq.~(\ref{eq:sigma_intermed_3})}

Noting that 
\begin{align}\label{eq:aside_int1}
\bm{G}(\tau_1-\tau'_2)\cdot\bm{G}(\tau_2-\tau'_2)&=\exp[-\bm{D}\cdot\bm{A}\left(\tau_{1}-\tau'_{2}\right)]\exp[-\bm{D}\cdot\bm{A}\left(\tau_{2}-\tau'_{2}\right)]\nonumber\\[10pt]
&=\exp[-\bm{D}\cdot\bm{A}\left(\tau_{1}+\tau_{2}\right)]\exp[2\bm{D}\cdot\bm{A}\tau'_2],
\end{align}
we may write
\begin{align}\label{eq:aside_int2}
\int_{0}^{\tau_2}d\tau'_{2}\exp[2\bm{D}\cdot\bm{A}\tau'_2]&=\dfrac{1}{2}\bm{D}^{-1}\cdot\bm{A}^{-1}\exp[2\bm{D}\cdot\bm{A}\tau'_2]\bigg\rvert_{0}^{\tau_2}\nonumber\\[10pt]
&=\dfrac{1}{2}\bm{D}^{-1}\cdot\bm{A}^{-1}\exp[2\bm{D}\cdot\bm{A}\tau_2]-\dfrac{1}{2}\bm{D}^{-1}\cdot\bm{A}^{-1},
\end{align}
which can be used to obtain the identity given in eq.~(\ref{eq:aside_int3}).

\subsection{Simplifying the RHS of eq.~(\ref{eq:wexp2}) to obtain an expression for $\left<W_{\text{dis}}\right>$ in the constant velocity pulling protocol}

The inner-most integral on the RHS of eq.~(\ref{eq:wexp2}) may be processed as follows. Recognizing that 
\begin{align}\label{eq:g_i_1}
\dfrac{d}{d\tau'}\bm{A}^{-1}\cdot\bm{D}^{-1}\cdot\bm{G}\left(\tau-\tau'\right)\cdot\bm{A}^{-1}=\bm{G}\left(\tau-\tau'\right)\cdot\bm{A}^{-1},
\end{align}
we may write
\begin{equation}\label{eq:g_intermed}
\begin{split}
\int_{0}^{\tau}d\tau'\left[\bm{G}(\tau-\tau')\cdot\bm{A}^{-1}\right]_{NN}&=\int_{0}^{\tau}d\tau'\left\{\dfrac{d}{d\tau'}\bm{A}^{-1}\cdot\bm{D}^{-1}\cdot\bm{G}\left(\tau-\tau'\right)\cdot\bm{A}^{-1}\right\}_{NN}\\[10pt]
&=\left[\bm{A}^{-1}\cdot\bm{D}^{-1}\cdot\bm{A}^{-1}-\bm{A}^{-1}\cdot\bm{D}^{-1}\cdot\bm{G}\left(\tau\right)\cdot\bm{A}^{-1}\right]_{NN}
\end{split}
\end{equation}
Plugging eq.~(\ref{eq:g_intermed}) into eq.~(\ref{eq:wexp2}), we get
\begin{align}\label{eq:g_i_2}
\left<W_{\text{dis}}\right>&=c^2v^2\int_{0}^{\tau_{\text{m}}}d\tau\Biggl[\bm{A}^{-1}\cdot\bm{D}^{-1}\cdot\bm{A}^{-1}-\bm{A}^{-1}\cdot\bm{D}^{-1}\cdot\bm{G}\left(\tau\right)\cdot\bm{A}^{-1}\Biggr]_{NN}\nonumber\\
&=c^2v^2\tau_{\text{m}}\left[\bm{A}^{-1}\cdot\bm{D}^{-1}\cdot\bm{A}^{-1}\right]_{NN}-c^2v^2\int_{0}^{\tau_{\text{m}}}d\tau\left[\bm{A}^{-1}\cdot\bm{D}^{-1}\cdot\bm{G}\left(\tau\right)\cdot\bm{A}^{-1}\right]_{NN}
\end{align}
Recognizing that
\begin{align}\label{eq:g_i_3}
\int_{0}^{\tau_{\text{m}}}d\tau\left[\bm{A}^{-1}\cdot\bm{D}^{-1}\cdot\bm{G}\left(\tau\right)\cdot\bm{A}^{-1}\right]_{NN}&=-\Biggl[\bm{A}^{-1}\cdot\bm{D}^{-1}\cdot\left[\exp\left(-\bm{D}\cdot\bm{A}\tau_{\text{m}}\right)-\boldsymbol{\delta}\right]\cdot\bm{A}^{-1}\cdot\bm{D}^{-1}\cdot\bm{A}^{-1}\Biggr]_{NN}
\end{align}
and combining eqs.~(\ref{eq:g_i_2}) and ~(\ref{eq:g_i_3}), one obtains eq.~(\ref{eq:wdis_const_vel}) in the main text.

\subsection{Simplifying the RHS of eq.~(\ref{eq:wdis_int_oscil}) to obtain an expression for $\left<W_{\text{dis}}\right>$ in the oscillatory driving protocol}

The innermost integral is evaluated as follows
\begin{align}\label{eq:os_1}
\int_{0}^{\tau}d\tau'\left[\cos(\omega\tau')\bm{G}(\tau-\tau')\cdot\bm{A}^{-1}\right]_{NN}&=\dfrac{1}{\omega}\left[\sin(\omega\tau')\bm{G}(\tau-\tau')\cdot\bm{A}^{-1}\right]_{NN}\Bigg\vert_{0}^{\tau}\nonumber\\[5pt]
&-\dfrac{1}{\omega}\int_{0}^{\tau}d\tau'\left[\sin(\omega\tau')\left(\bm{D}\cdot\bm{A}\right)\cdot\bm{G}(\tau-\tau')\cdot\bm{A}^{-1}\right]_{NN}
\end{align}
Processing the second integral on the RHS of the above equation by parts, 
\begin{align}\label{eq:os_2}
\int_{0}^{\tau}d\tau'\left[\sin(\omega\tau')\left(\bm{D}\cdot\bm{A}\right)\cdot\bm{G}(\tau-\tau')\cdot\bm{A}^{-1}\right]_{NN}&=-\dfrac{1}{\omega}\left[\cos(\omega\tau')\left(\bm{D}\cdot\bm{A}\right)\cdot\bm{G}(\tau-\tau')\cdot\bm{A}^{-1}\right]_{NN}\Bigg\vert_{0}^{\tau}\nonumber\\[5pt]
&+\dfrac{1}{\omega}\int_{0}^{\tau}d\tau'\left[\cos(\omega\tau')\left(\bm{D}\cdot\bm{A}\right)^2\cdot\bm{G}(\tau-\tau')\cdot\bm{A}^{-1}\right]_{NN}
\end{align}
Plugging eq.~(\ref{eq:os_2}) into eq.~(\ref{eq:os_1}) and simplifying, 
\begin{align}\label{eq:os_4}
\int_{0}^{\tau}d\tau'\left[\cos(\omega\tau')\left\{\omega^2\boldsymbol{\delta}+\left(\bm{D}\cdot\bm{A}\right)^2\right\}\cdot\bm{G}(\tau-\tau')\cdot\bm{A}^{-1}\right]_{NN}&=\left[\omega\sin(\omega\tau)\bm{A}^{-1}\right]_{NN}\nonumber\\[5pt]
&+\left[\cos(\omega\tau')\left(\bm{D}\cdot\bm{A}\right)\cdot\bm{G}(\tau-\tau')\cdot\bm{A}^{-1}\right]_{NN}\Bigg\vert_{0}^{\tau}
\end{align}
The above equality holds for not just the $(NN)^{\text{th}}$ matrix element, but also for the matrices themselves. Therefore, we may write
\begin{align}\label{eq:os_5}
\int_{0}^{\tau}d\tau'\left[\cos(\omega\tau')\left\{\omega^2\boldsymbol{\delta}+\left(\bm{D}\cdot\bm{A}\right)^2\right\}\cdot\bm{G}(\tau-\tau')\cdot\bm{A}^{-1}\right]&=
\left[\omega\sin(\omega\tau)\bm{A}^{-1}\right]\nonumber\\[5pt]
&+\left[\cos(\omega\tau')\bm{G}(\tau-\tau')\cdot\left(\bm{D}\cdot\bm{A}\right)\cdot\bm{A}^{-1}\right]\Bigg\vert_{0}^{\tau}
\end{align}
Defining 
\begin{align}\label{eq:phi_def}
\boldsymbol{\Phi}=\left[\omega^2\boldsymbol{\delta}+\left(\bm{D}\cdot\bm{A}\right)^2\right]^{-1}
\end{align}
and premultiplying by $\boldsymbol{\Phi}$ on both sides of eq.~(\ref{eq:os_5}), yields, after simplification,
\begin{align}\label{eq:os_7}
&\int_{0}^{\tau}d\tau'\left[\cos(\omega\tau')\bm{G}(\tau-\tau')\cdot\bm{A}^{-1}\right]=\Bigl[\omega\sin(\omega\tau)\boldsymbol{\Phi}\cdot\bm{A}^{-1}+\cos(\omega\tau)\boldsymbol{\Phi}\cdot\bm{D}-\boldsymbol{\Phi}\cdot\bm{G}(\tau)\cdot\bm{D}\Bigr]
\end{align}
Since the above matrix relation also holds true for the $(NN)^{\text{th}}$ element, we therefore have
\begin{align}\label{eq:inner_intg_oscil}
&\int_{0}^{\tau}d\tau'\left[\cos(\omega\tau')\bm{G}(\tau-\tau')\cdot\bm{A}^{-1}\right]_{NN}=\Bigl[\omega\sin(\omega\tau)\boldsymbol{\Phi}\cdot\bm{A}^{-1}+\cos(\omega\tau)\boldsymbol{\Phi}\cdot\bm{D}-\boldsymbol{\Phi}\cdot\bm{G}(\tau)\cdot\bm{D}\Bigr]_{NN}
\end{align}
Substituting eq.~(\ref{eq:inner_intg_oscil}) into eq.~(\ref{eq:wdis_int_oscil}), we obtain
\begin{align}\label{eq:int_wdis_o}
\begin{split}
\left<W_{\text{dis}}\right>&=c^2d^2\omega^2\int_{0}^{\tau_{\text{m}}}d\tau\left[\omega\cos(\omega\tau)\sin(\omega\tau)\boldsymbol{\Phi}\cdot\bm{A}^{-1}\right]_{NN}+c^2d^2\omega^2\int_{0}^{\tau_{\text{m}}}d\tau\left[\cos^2(\omega\tau)\boldsymbol{\Phi}\cdot\bm{D}\right]_{NN}\\[5pt]
&-c^2d^2\omega^2\int_{0}^{\tau_{\text{m}}}d\tau\left[\cos(\omega\tau)\boldsymbol{\Phi}\cdot\bm{G}(\tau)\cdot\bm{D}\right]_{NN}
\end{split}
\end{align}
The first two integrals on the RHS of the above equation are easily evaluated, using the double-angle formulae $\cos(\omega\tau)\sin(\omega\tau)=(1/2)[\sin(2\omega\tau)]$ and $\cos^2(\omega\tau)=(1/2)[1+\cos(2\omega\tau)]$, respectively, to give
\begin{align}\label{eq:int1_os}
\int_{0}^{\tau_{\text{m}}}d\tau\left[\omega\cos(\omega\tau)\sin(\omega\tau)\boldsymbol{\Phi}\cdot\bm{A}^{-1}\right]_{NN}=\dfrac{1}{4}\left[1-\cos(2\omega\tau_{\text{m}})\right]\left[\boldsymbol{\Phi}\cdot\bm{A}^{-1}\right]_{NN}
\end{align}
\begin{align}\label{eq:int2_os}
\int_{0}^{\tau_{\text{m}}}d\tau\left[\cos^2(\omega\tau)\boldsymbol{\Phi}\cdot\bm{D}\right]_{NN}=\left[\dfrac{\tau_{\text{m}}}{2}+\dfrac{\sin\left(2\omega\tau_{\text{m}}\right)}{4\omega}\right]\left[\boldsymbol{\Phi}\cdot\bm{D}\right]_{NN}
\end{align}
The last integral on the RHS of eq.~(\ref{eq:int_wdis_o}) is evaluated by parts, as follows
\begin{align}\label{eq:m1}
&\int_{0}^{\tau_{\text{m}}}d\tau\left[\cos(\omega\tau)\boldsymbol{\Phi}\cdot\bm{G}(\tau)\cdot\bm{D}\right]_{NN}=\dfrac{1}{\omega}\left[\sin(\omega\tau)\boldsymbol{\Phi}\cdot\bm{G}(\tau)\cdot\bm{D}\right]_{NN}\Bigg\vert_{0}^{\tau_{\text{m}}}\nonumber\\[5pt]
&-\dfrac{1}{\omega}\int_{0}^{\tau_{\text{m}}}d\tau\left[\sin(\omega\tau)\boldsymbol{\Phi}\cdot\left(\bm{D}\cdot\bm{A}\right)\cdot\bm{G}(\tau)\cdot\bm{D}\right]_{NN}
\end{align}
Processing the second integral on the RHS of the above equation by parts, and simplifying eq.~(\ref{eq:m1}) using a procedure identical to that adopted in eqs.~(\ref{eq:os_4})-(\ref{eq:inner_intg_oscil}), we get
\begin{align}\label{eq:outer_intg_oscil}
\int_{0}^{\tau_{\text{m}}}d\tau\left[\cos(\omega\tau)\boldsymbol{\Phi}\cdot\bm{G}(\tau)\cdot\bm{D}\right]_{NN}=
\Biggl[\omega\sin(\omega\tau_{\text{m}})\boldsymbol{\Phi}^2\cdot\bm{G}(\tau_{\text{m}})\cdot\bm{D}
+\cos(\omega\tau_{\text{m}})\boldsymbol{\Phi}^2\cdot\left(\bm{D}\cdot\bm{A}\right)\cdot\bm{G}(\tau_{\text{m}})\cdot\bm{D}-\boldsymbol{\Phi}^2\cdot\bm{D}\cdot\bm{A}\cdot\bm{D}\Biggr]_{NN}
\end{align}

Substituting eqs.~(\ref{eq:int1_os}), ~(\ref{eq:int2_os}), and ~(\ref{eq:outer_intg_oscil}) into eq.~(\ref{eq:int_wdis_o}) and simplifying, one obtains eq.~(\ref{eq:wdis_o_fin}).

\subsection{Deriving an expression for $\left<W_{\text{dis}}\right>$ in the oscillatory driving protocol, for a single-mode bead-spring dashpot chain ($N=1$)}

For a dumbbell ($N=1$) subjected to oscillatory driving, eq.~(\ref{eq:n1_sim}) applies, with $c_2=c$, along with
\begin{equation}\label{eq:n1_phi}
\begin{split}
\left[\boldsymbol{\Phi}\right]_{NN}&=\dfrac{\left(1+\varphi\right)^2}{\omega^2\left(1+\varphi\right)^2+(c+1)^2}\\[5pt]
\left[\boldsymbol{\Phi}\cdot\bm{A}^{-1}\right]_{NN}&=\left(\dfrac{1}{c+1}\right)\left[\dfrac{\left(1+\varphi\right)^2}{\omega^2\left(1+\varphi\right)^2+(c+1)^2}\right]\\[5pt]
\left[\boldsymbol{\Phi}\cdot\bm{D}\right]_{NN}&=\left[\dfrac{\left(1+\varphi\right)}{\omega^2\left(1+\varphi\right)^2+(c+1)^2}\right]\\[5pt]
\left[\boldsymbol{\Phi}^2\cdot\bm{G}(\tau_{\text{m}})\cdot\bm{D}\right]_{NN}&=\dfrac{\left(1+\varphi\right)^3}{\left[\omega^2\left(1+\varphi\right)^2+(c+1)^2\right]^2}\exp\left[-\dfrac{c+1}{1+\varphi}\tau_{\text{m}}\right]\\[5pt]
\left[\boldsymbol{\Phi}^2\cdot\left(\bm{D}\cdot\bm{A}\right)\cdot\bm{G}(\tau_{\text{m}})\cdot\bm{D}\right]_{NN}&=\dfrac{\left(1+\varphi\right)^2(c+1)}{\left[\omega^2\left(1+\varphi\right)^2+(c+1)^2\right]^2}\exp\left[-\dfrac{c+1}{1+\varphi}\tau_{\text{m}}\right]\\[5pt]
\left[\boldsymbol{\Phi}\cdot\bm{D}\cdot\bm{A}\cdot\bm{D}\right]_{NN}&=\dfrac{\left(1+\varphi\right)^2(c+1)}{\left[\omega^2\left(1+\varphi\right)^2+(c+1)^2\right]^2}
\end{split}
\end{equation}
Substituting eq.~(\ref{eq:n1_phi}) into eq.~(\ref{eq:wdis_o_num}) and simplifying, eq.~(\ref{eq:n1_wdis_os}) is obtained.

\end{widetext}

\bibliography{ms_rev_2}

@Book{Callen1985,
  Title                    = {Thermodynamics and an Introduction to Thermostatics},
  Author                   = {H. B. Callen},
  Publisher                = {John Wiley and Sons, New York},
  Year                     = {1985}
}

@Book{Bird1987b,
  Title                    = {Dynamics of Polymeric Liquids - Volume 2 : Kinetic Theory},
  Author                   = {R. B. Bird and C. F. Curtiss and R. C. Armstrong and O. Hassager},
  Publisher                = {John Wiley and Sons, New York},
  Year                     = {1987}
   }

@Book{Ottinger1996,
  Title                    = {Stochastic Processes in Polymeric Fluids},
  Author                   = {H. C. \"{O}ttinger},
  Publisher                = {Springer, Berlin},
  Year                     = {1996}
}

@Book{degennes,
  Title                    = {Scaling Concepts in Polymer Physics},
  Author                   = {P.-G. {de G}ennes},
  Publisher                = {Cornell University Press, Ithaca},
  Year                     = {1979}
 }

@InCollection{ravibook,
  author    = {Prakash, J R},
  booktitle = {Rheology Series},
  publisher = {Elsevier},
  title     = {{The kinetic theory of dilute solutions of flexible polymers: Hydrodynamic interaction}},
  year      = {1999},
  editor    = {Siginer, D A and De Kee, D and Chhabra, R P},
  pages     = {467--517},
  volume    = {8},
}

@article{kuhn1945,
author = {Kuhn, Werner and Kuhn, Hans},
journal = {Helv. Chim. Acta},
number = {1},
pages = {1533--1579},
publisher = {Wiley Online Library},
title = {{Bedeutung beschr{\"{a}}nkt freier Drehbarkeit f{\"{u}}r die Viskosit{\"{a}}t und Str{\"{o}}mungsdoppelbrechung von Fadenmolekell{\"{o}}sungen I}},
volume = {28},
year = {1945}
}

@article{Massa1971,
author = {Massa, Dennis J. and Schrag, John L. and Ferry, John D.},
issn = {15205835},
journal = {Macromolecules},
number = {2},
pages = {210--214},
title = {{Dynamic Viscoelastic Properties of Polystyrene in High-Viscosity Solvents. Extrapolation to Infinite Dilution and High-Frequency Behavior}},
volume = {4},
year = {1971}
}

@article{Mackay1992,
author = {Mackay, M. E. and Liang, C. H. and Halley, P. J.},
journal = {Rheol. Acta},
number = {5},
pages = {481--489},
publisher = {Steinkopff-Verlag},
title = {{Instrument effects on stress jump measurements}},
volume = {31},
year = {1992}
}

@article{Manke1987,
author = {Manke, Charles W and Williams, Michael C},
issn = {15208516},
journal = {J. Rheol.},
number = {6},
pages = {495--510},
title = {{Stress Jump at the Inception of Shear and Elongational Flows of Dilute Polymer Solutions, Due to Internal Viscosity}},
volume = {31},
year = {1987}
}

@article{Khatri20076770,
author = {Khatri, B S and McLeish, T C B},
journal = {Macromolecules},
number = {18},
pages = {6770--6777},
title = {{Rouse model with internal friction: A coarse grained framework for single biopolymer dynamics}},
volume = {40},
year = {2007}
}

@article{Khatri20071825,
author = {Khatri, B S and Kawakami, M and Byrne, K and Smith, D A and McLeish, T C B},
journal = {Biophys. J.},
number = {6},
pages = {1825--1835},
title = {{Entropy and barrier-controlled fluctuations determine conformational viscoelasticity of single biomolecules}},
volume = {92},
year = {2007}
}

@article{Poirier2002,
author = {Poirier, Michael G. and Marko, John F.},
journal = {Phys. Rev. Lett.},
number = {22},
pages = {228103},
title = {{Effect of Internal Friction on Biofilament Dynamics}},
volume = {88},
year = {2002}
}

@article{Mondal2020,
author = {Mondal, Debasmita and Adhikari, Ronojoy and Sharma, Prerna},
journal = {Sci. Adv.},
number = {33},
pages = {eabb0503},
title = {{Internal friction controls active ciliary oscillations near the instability threshold}},
volume = {6},
year = {2020}
}

@article{Murayama2007,
author = {Murayama, Y. and Wada, H. and Sano, M.},
isbn = {0295-5075},
issn = {02955075},
journal = {Eur. Phys. Lett.},
number = {5},
title = {{Dynamic force spectroscopy of a single condensed DNA}},
volume = {79},
pages = {58001},
year = {2007}
}

@article{ja211494h,
author = {Schulz, Julius C F and Schmidt, Lennart and Best, Robert B and Dzubiella, Joachim and Netz, Roland R},
journal = {J. Am. Chem. Soc.},
number = {14},
pages = {6273--6279},
title = {{Peptide Chain Dynamics in Light and Heavy Water: Zooming in on Internal Friction}},
volume = {134},
year = {2012}
}

@article{Soranno2012,
author = {Soranno, A and Buchli, B and Nettels, D and Cheng, R R and M{\"{u}}ller-Sp{\"{a}}th, S and Pfeil, S H and Hoffmann, A and Lipman, E A and Makarov, D E and Schuler, B},
journal = {Proc. Natl. Acad. Sci. U.S.A.},
number = {44},
pages = {17800--17806},
title = {{Quantifying internal friction in unfolded and intrinsically disordered proteins with single-molecule spectroscopy}},
volume = {109},
year = {2012}
}

@article{Soranno2017,
author = {Soranno, Andrea and Holla, Andrea and Dingfelder, Fabian and Nettels, Daniel and Makarov, Dmitrii E and Schuler, Benjamin},
journal = {Proc. Natl. Acad. Sci. U.S.A.},
number = {10},
pages = {E1833--E1839},
title = {{Integrated view of internal friction in unfolded proteins from single-molecule FRET, contact quenching, theory, and simulations}},
volume = {114},
year = {2017}
}

@article{Alexander-Katz2009,
author = {Alexander-Katz, A and Wada, H and Netz, R R},
journal = {Phys. Rev. Lett.},
number = {2},
title = {{Internal friction and nonequilibrium unfolding of polymeric globules}},
volume = {103},
pages={028102},
year = {2009}
}

@article{Einert2011,
author = {Einert, T. R. and Sing, C. E. and Alexander-Katz, A. and Netz, R. R.},
journal = {Eur. Phys. J. E},
number = {12},
pages = {130},
pmid = {22167584},
title = {{Conformational dynamics and internal friction in homopolymer globules: Equilibrium vs. non-equilibrium simulations}},
volume = {34},
year = {2011}
}

@article{Samanta2016,
author = {Samanta, Nairhita and Chakrabarti, Rajarshi},
issn = {03784371},
journal = {Physica A},
pages = {165--179},
publisher = {Elsevier B.V.},
title = {{Reconfiguration dynamics in folded and intrinsically disordered protein with internal friction: Effect of solvent quality and denaturant}},
volume = {450},
year = {2016}
}

@article{Kailasham2018,
author = {Kailasham, R. and Chakrabarti, Rajarshi and Prakash, J. Ravi},
journal = {J. Chem. Phys.},
number = {9},
pages = {094903},
title = {{Rheological consequences of wet and dry friction in a dumbbell model with hydrodynamic interactions and internal viscosity}},
volume = {149},
year = {2018}
}

@article{Kailasham2020,
author = {Kailasham, R. and Chakrabarti, Rajarshi and Prakash, J. Ravi},
journal = {Phys. Rev. Res.},
pages = {013331},
title = {{Wet and dry internal friction can be measured with the Jarzynski equality}},
volume = {2},
year = {2020}
}

@article{kailasham2021rouse,
author = {Kailasham, R. and Chakrabarti, Rajarshi and Prakash, J. Ravi},
journal = {J. Rheol.},
pages = {903},
title = {{Rouse model with fluctuating internal friction}},
volume = {65},
year = {2021}
}

@article{Kailasham2023,
author = {Kailasham, R and Chakrabarti, Rajarshi and Prakash, J Ravi},
journal = {J. Rheol.},
number = {1},
pages = {105--123},
publisher = {The Society of Rheology},
title = {{Shear viscosity for finitely extensible chains with fluctuating internal friction and hydrodynamic interactions}},
volume = {67},
year = {2023}
}

@article {Nandagiri2020,
title = {Flagellar energetics from high-resolution imaging of beating patterns in tethered mouse sperm},
author = {Nandagiri, Ashwin and Gaikwad, Avinash Satish and Potter, David L and Nosrati, Reza and Soria, Julio and O'Bryan, Moira K and Jadhav, Sameer and Prabhakar, Ranganathan},
volume = 10,
year = 2021,
pages = {e62524},
doi = {10.7554/eLife.62524},
url = {https://doi.org/10.7554/eLife.62524},
journal = {eLife},
issn = {2050-084X},
publisher = {eLife Sciences Publications, Ltd},
}

@article{Dasbach1992,
author = {Dasbach, T. P. and Manke, C. W. and Williams, M. C.},
journal = {J. Phys. Chem.},
number = {10},
pages = {4118--4125},
title = {{Complex viscosity for the rigorous formulation of the multibead internal viscosity model with hydrodynamic interaction}},
volume = {96},
year = {1992}
}

@article{Booij1970,
author = {Booij, H. C. and van Wiechen, P. H.},
journal = {J. Chem. Phys.},
number = {10},
pages = {5056--5068},
title = {{Effect of Internal Viscosity on the Deformation of a Linear Macromolecule in a Sheared Solution}},
volume = {52},
year = {1970}
}

@article{Hua1995,
author = {Hua, C C and Schieber, J D},
journal = {J. Non-Newton. Fluid Mech.},
number = {3},
pages = {307--332},
title = {{Nonequilibrium Brownian dynamics simulations of Hookean and FENE dumbbells with internal viscosity}},
volume = {56},
year = {1995}
}

@article{Hua1996,
author = {Hua, Chi C. and Schieber, Jay D. and Manke, Charles W.},
issn = {0035-4511},
journal = {Rheol. Acta},
number = {3},
pages = {225--232},
title = {{Linear viscoelastic behavior of the Hookean dumbbell with internal viscosity}},
volume = {35},
year = {1996}
}

@article{Hagen2010385,
author = {Hagen, S J},
journal = {Curr. Protein Pept. Sci.},
number = {5},
pages = {385--395},
title = {{Solvent viscosity and friction in protein folding dynamics}},
volume = {11},
year = {2010}
}

@article{Ansari1992,
author = {Ansari, Anjum and Jones, Colleen M and Henry, Eric R and Hofrichter, James and Eaton, William A},
journal = {Science},
number = {5065},
pages = {1796--1798},
title = {{The Role of Solvent Viscosity in the Dynamics of Protein Conformational Changes}},
volume = {256},
year = {1992}
}

@article{Qiu2004,
author = {Qiu, L and Hagen, S J},
journal = {J. Am. Chem. Soc.},
number = {11},
pages = {3398--3399},
title = {{A Limiting Speed for Protein Folding at Low Solvent Viscosity}},
volume = {126},
year = {2004}
}

@article{Wensley2010,
author = {Wensley, Beth G and Batey, Sarah and Bone, Fleur A C and Chan, Zheng Ming and Tumelty, Nuala R and Steward, Annette and Kwa, Lee Gyan and Borgia, Alessandro and Clarke, Jane},
isbn = {1089-8638 (Electronic)$\backslash$n0022-2836 (Linking)},
issn = {0028-0836},
journal = {Nature},
number = {7281},
pages = {685--688},
pmid = {20130652},
publisher = {Nature Publishing Group},
title = {{Experimental evidence for a frustrated energy landscape in a three-helix-bundle protein family}},
volume = {463},
year = {2010}
}

@article{Schulz20154565,
author = {Schulz, J C F and Miettinen, M S and Netz, R R},
journal = {J. Phys. Chem. B},
number = {13},
pages = {4565--4574},
title = {{Unfolding and Folding Internal Friction of $\beta$-Hairpins Is Smaller than That of $\alpha$-Helices}},
volume = {119},
year = {2015}
}

@article{Cellmer2008,
author = {Cellmer, T. and Henry, E. R. and Hofrichter, J. and Eaton, W. A.},
isbn = {1091-6490 (Electronic)$\backslash$r0027-8424 (Linking)},
issn = {0027-8424},
journal = {Proc. Natl. Acad. Sci. U.S.A.},
number = {47},
pages = {18320--18325},
title = {{Measuring internal friction of an ultrafast-folding protein}},
volume = {105},
year = {2008}
}

@article{Dhar2005,
author = {Dhar, Abhishek},
journal = {Phys. Rev. E},
number = {3},
pages = {036126},
primaryClass = {cond-mat},
title = {Work distribution functions in polymer stretching experiments},
volume = {71},
year = {2005}
}

@article{DeSancho2014,
author = {de Sancho, David and Sirur, Anshul and Best, Robert B},
isbn = {2122633255},
issn = {2041-1723},
journal = {Nat. Commun.},
pages = {4307},
pmid = {24986114},
publisher = {Nature Publishing Group},
title = {{Molecular origins of internal friction effects on protein-folding rates.}},
volume = {5},
year = {2014}
}

@article{Echeverria2014,
author = {Echeverria, Ignacia and Makarov, Dmitrii E. and Papoian, Garegin A.},
issn = {15205126},
journal = {J. Am. Chem. Soc.},
number = {24},
pages = {8708--8713},
pmid = {24844314},
title = {{Concerted dihedral rotations give rise to internal friction in unfolded proteins}},
volume = {136},
year = {2014}
}

@article{Ameseder2018,
author = {Ameseder, Felix and Radulescu, Aurel and Holderer, Olaf and Falus, Peter and Richter, Dieter and Stadler, Andreas M.},
journal = {J. Phys. Chem. Lett.},
pages = {2469--2473},
title = {{Relevance of Internal Friction and Structural Constraints for the Dynamics of Denatured Bovine Serum Albumin}},
volume = {9},
year = {2018}
}

@article{Daldrop2018,
author = {Daldrop, Jan O. and Kappler, Julian and Brunig, Florian N. and Netz, Roland R},
journal = {Proc. Natl. Acad. Sci. U.S.A.},
number = {20},
pages = {5169--5174},
title = {{Butane dihedral angle dynamics in water is dominated by internal friction}},
volume = {115},
year = {2018}
}

@article{Jas2001,
author = {Jas, Gouri S. and Eaton, William A. and Hofrichter, James},
isbn = {1520-6106},
issn = {1520-6106},
journal = {J. Phys. Chem. B},
number = {1},
pages = {261--272},
pmid = {166324400038},
title = {{Effect of Viscosity on the Kinetics of $\alpha$-Helix and $\beta$-Hairpin Formation}},
volume = {105},
year = {2001}
}

@article{Jarzynski1997,
author = {Jarzynski, C.},
journal = {Phys. Rev. Lett.},
number = {14},
pages = {2690--2693},
title = {{Nonequilibrium Equality for Free Energy Differences}},
volume = {78},
year = {1997}
}

@article{Speck2004,
author = {Speck, Thomas and Seifert, Udo},
journal = {Phys. Rev. E},
number = {6},
pages = {066112},
title = {{Distribution of work in isothermal nonequilibrium processes}},
volume = {70},
year = {2004}
}

@article{Speck2005,
author = {Speck, T. and Seifert, U.},
isbn = {1434-6028},
issn = {14346028},
journal = {Eur. Phys. J. B},
number = {4},
pages = {521--527},
pmid = {11590329},
title = {{Dissipated work in driven harmonic diffusive systems: General solution and application to stretching Rouse polymers}},
volume = {43},
year = {2005}
}

@article{Varghese2013,
author = {Varghese, Anoop and Vemparala, Satyavani and Rajesh, R.},
issn = {15393755},
journal = {Phys. Rev. E},
number = {2},
pages = {022134},
title = {{Force fluctuations in stretching a tethered polymer}},
volume = {88},
year = {2013}
}

@article{Harris2007,
author = {Harris, N. C. and Song, Y. and Kiang, C.-H.},
issn = {0031-9007},
journal = {Phys. Rev. Lett.},
number = {6},
pages = {068101},
pmid = {17930869},
title = {{Experimental Free Energy Surface Reconstruction from Single-Molecule Force Spectroscopy using Jarzynski's Equality}},
volume = {99},
year = {2007}
}

@article{Latinwo2014,
author = {Latinwo, Folarin and Hsiao, Kai-Wen and Schroeder, Charles M.},
issn = {0021-9606},
journal = {J. Chem. Phys.},
number = {17},
pages = {174903},
pmid = {25381543},
title = {{Nonequilibrium thermodynamics of dilute polymer solutions in flow}},
volume = {141},
year = {2014}
}

@article{Latinwo2013,
author = {Latinwo, Folarin and Schroeder, Charles M.},
journal = {Macromolecules},
number = {20},
pages = {8345--8355},
title = {{Nonequilibrium work relations for polymer dynamics in dilute solutions}},
volume = {46},
year = {2013}
}

@article{Latinwo2014a,
author = {Latinwo, Folarin and Schroeder, Charles M.},
journal = {Soft matter},
number = {13},
pages = {2178--87},
title = {{Determining elasticity from single polymer dynamics.}},
volume = {10},
year = {2014}
}

@article{Zhou2022,
author = {Zhou, Yuecheng and Latinwo, Folarin and Schroeder, Charles M.},
journal = {Entropy},
number = {1},
pages = {27},
title = {{Crooks fluctuation theorem for single polymer dynamics in time-dependent flows: Understanding viscoelastic hysteresis}},
volume = {24},
year = {2022}
}

@article{Trepagnier2004,
author = {Trepagnier, E. H. and Jarzynski, C. and Ritort, F. and Crooks, G. E. and Bustamante, C. J. and Liphardt, J.},
isbn = {0027-8424 (Print)},
issn = {0027-8424},
journal = {Proc. Natl. Acad. Sci. U.S.A.},
number = {42},
pages = {15038--15041},
pmid = {15469914},
title = {{Experimental test of Hatano and Sasa's nonequilibrium steady-state equality}},
volume = {101},
year = {2004}
}

@article{Liphardt2002,
author = {Liphardt, Jan and Dumont, Sophie and Smith, Steven B. and {Tinoco Jr.}, Ignacio and Bustamante, Carlos},
isbn = {1095-9203 (Electronic)$\backslash$r0036-8075 (Linking)},
issn = {00368075},
journal = {Science},
number = {5574},
pages = {1832--1835},
pmid = {12052949},
title = {{Equilibrium information from nonequilibrium measurements in an experimental test of Jarzynski's equality}},
volume = {296},
year = {2002}
}

@article{Fixman1988,
author = {Fixman, M.},
issn = {00219606},
journal = {J. Chem. Phys.},
number = {4},
pages = {2442},
title = {{Dynamics of stiff polymer chains}},
volume = {89},
year = {1988}
}

@article{Ghosal2016,
author = {Ghosal, Aishani and Cherayil, Binny J.},
journal = {J. Chem. Phys.},
number = {20},
pages = {204901},
title = {{Polymer extension under flow: Some statistical properties of the work distribution function}},
volume = {145},
year = {2016}
}

@article{Sharma2011,
author = {Sharma, Rati and Cherayil, Binny J.},
journal = {Phys. Rev. E},
number = {4},
pages = {041805},
title = {{Work fluctuations in an elastic dumbbell model of polymers in planar elongational flow}},
volume = {83},
year = {2011}
}

@article{Ghosal2016a,
author = {Ghosal, Aishani and Cherayil, Binny J.},
journal = {J. Chem. Phys.},
number = {21},
title = {{Polymer extension under flow: A path integral evaluation of the free energy change using the Jarzynski relation}},
volume = {144},
year = {2016}
}

@article{Swiatek2024,
author = {{\'{S}}wi{\c{a}}tek, Adam and Kuczera, Krzysztof and Szoszkiewicz, Robert},
issn = {15205207},
journal = {J. Phys. Chem. B},
number = {16},
pages = {3856--3869},
title = {{Effects of Proline on Internal Friction in Simulated Folding Dynamics of Several Alanine-Based $\alpha$-Helical Peptides}},
volume = {128},
year = {2024}
}

@article{Ojala2014,
author = {Ojala, Heikki and Ziedaite, Gabija and Wallin, Anders E. and Bamford, Dennis H. and H{\ae}ggstr{\"{o}}m, Edward},
journal = {Eur. Biophys. J.},
number = {2-3},
pages = {71--79},
title = {{Optical tweezers reveal force plateau and internal friction in PEG-induced DNA condensation}},
volume = {43},
year = {2014}
}

@article{Braun2004a,
author = {Braun, Oliver and Hanke, Andreas and Seifert, Udo},
issn = {00319007},
journal = {Phys. Rev. Lett.},
number = {15},
pages = {158105},
title = {{Probing molecular free energy landscapes by periodic loading}},
volume = {93},
year = {2004}
}

@article{Woodside2014,
author = {Woodside, Michael T. and Block, Steven M.},
journal = {Ann. Rev. Biophys.},
number = {1},
pages = {19--39},
pmid = {24895850},
title = {{Reconstructing Folding Energy Landscapes by Single-Molecule Force Spectroscopy}},
volume = {43},
year = {2014}
}

@article{Szymczak2009,
author = {Szymczak, Piotr and Janovjak, Harald},
journal = {J. Mol. Biol.},
number = {3},
pages = {443--456},
title = {{Periodic Forces Trigger a Complex Mechanical Response in Ubiquitin}},
volume = {390},
year = {2009}
}

@article{Wu2018,
author = {Wu, Meiling and Lu, H. Peter},
journal = {J. Phys. Chem. B},
number = {51},
pages = {12312--12321},
pmid = {30481025},
title = {{Oscillating Piconewton Force Manipulation on Single-Molecule Enzymatic Conformational and Reaction Dynamics}},
volume = {122},
year = {2018}
}

@article{Blaber2022,
author = {Blaber, Steven and Sivak, David A.},
journal = {Epl},
number = {1},
pages = {17001},
title = {{Efficient two-dimensional control of barrier crossing}},
volume = {139},
year = {2022}
}

@article{Blaber2022a,
author = {Blaber, Steven and Sivak, David A.},
journal = {Phys. Rev. E},
number = {2},
pages = {L022103},
pmid = {36110009},
title = {{Optimal control with a strong harmonic trap}},
volume = {106},
year = {2022}
}

@misc{Kailasham_MATLAB_2025,
author = {Kailasham, R.},
month = jul,
title = {{MATLAB codes for calculating dissipation in driven bead-spring-dashpot chains}},
note={https://github.com/rkailasham/bsp{\_}dashpot{\_}lin{\_}sym{\_}driving},
url = {https://github.com/rkailasham/bsp_dashpot_lin_sym_driving},
year = {2025}
}

@PhdThesis{rk2021,
  author       = {Kailasham, R.},
  title	       = {The Influence of Internal Friction on Dilute Polymer Solution Dynamics},
  school       = {IIT Bombay-Monash Research Academy},
  year	       = 2021,
  month	       = {August},
  url	       = {https://doi.org/10.26180/15138882}
}

@article{Piana2015,
author = {Piana, Stefano and Donchev, Alexander G and Robustelli, Paul and Shaw, David E},
journal = {J. Phys. Chem. B},
pages = {5113--5123},
title = {{Water Dispersion Interactions Strongly In fl uence Simulated Structural Properties of Disordered Protein States}},
volume = {119},
year = {2015}
}

@article{Toyabe2010,
author = {Toyabe, Shoichi and Sagawa, Takahiro and Ueda, Masahito and Muneyuki, Eiro and Sano, Masaki},
issn = {1745-2473},
journal = {Nat. Phys.},
number = {November},
pages = {998--992},
title = {{Experimental demonstration of information-to-energy conversion and validation of the generalized Jarzynski equality}},
volume = {6},
year = {2010}
}

@article{Hatano2001,
author = {Hatano, T. and Sasa, S.},
journal = {Phys. Rev. Lett.},
number = {16},
pages = {3463--3466},
title = {{Steady-state thermodynamics of Langevin systems}},
volume = {86},
year = {2001}
}

@article{Harada2005,
author = {Harada, T. and Sasa, S.},
journal = {Phys. Rev. Lett.},
number = {13},
pages = {130602},
title = {{Equality connecting energy dissipation with a violation of the fluctuation-response relation}},
volume = {95},
year = {2005}
}

\end{document}